\documentclass[12pt]{article}
\usepackage{amsfonts,amsthm,amsmath,amssymb,times}
\usepackage{fullpage,graphicx,multicol}
\usepackage{epic,eepic}

\usepackage[small]{caption}

\newcounter{kevinslistcounter}%
\renewenvironment{itemize}{
\begin{list}{(\alph{kevinslistcounter})}
{\usecounter{kevinslistcounter}
\setlength{\parsep}{0pt}
\setlength{\labelwidth}{24pt}
\setlength{\itemsep}{0pt}
\setlength{\topsep}{\parsep}}}{\end{list}}

\newcounter{kevinslistcountertoo}%
\renewenvironment{enumerate}{
\begin{list}{\arabic{kevinslistcountertoo}.}
{\usecounter{kevinslistcountertoo}
\setlength{\parsep}{0pt}
\setlength{\labelwidth}{24pt}
\setlength{\itemsep}{6pt}
\setlength{\topsep}{\parsep}}}{\end{list}}

\newcommand{\includegraphicsbw}[2]{\includegraphics[#2]{#1.eps}}
\renewcommand{\includegraphicsbw}[2]{\includegraphics[#2]{#1_bw.eps}}

\newtheorem{proposition}{Proposition}[section]

\newtheorem{theorem}{Theorem}
\newtheorem{lemma}{Lemma}[section]

\newcommand{\tht}{\theta}
\newcommand{\la}{\lambda}
\newcommand{\eps}{\varepsilon}
\newcommand{\aff}{a_{\rm ff}}
\newcommand{\afb}{a_{\rm fb}}

\newcommand{\bec}{\begin{center}}
\newcommand{\eec}{\end{center}}
\newcommand{\bei}{\begin{itemize}}
\newcommand{\eei}{\end{itemize}}
\newcommand{\beq}{\begin{equation}}
\newcommand{\eeq}{\end{equation}}
\newcommand{\beqn}{\begin{equation*}}
\newcommand{\eeqn}{\end{equation*}}
\newcommand{\beqr}{\begin{eqnarray}}
\newcommand{\eeqr}{\end{eqnarray}}
\newcommand{\beqrn}{\begin{eqnarray*}}
\newcommand{\eeqrn}{\end{eqnarray*}}
\newcommand{\brr}{\begin{array}}
\newcommand{\err}{\end{array}}
\newcommand{\bef}{\begin{figure}}
\newcommand{\eef}{\end{figure}}

\newcommand{\om}{\omega}

\newcommand{\R}{\mathbb R}
\newcommand{\T}{\mathbb T}

\newcommand{\lmax}{\lambda_{\rm max}}
\newcommand{\lmin}{\lambda_{\rm min}}

\newcommand{\delom}{\Delta \omega}

\newcommand{\delafb}{\Delta \afb}

\date{August 22, 2007}

\title{Reliability of Coupled Oscillators I:\\ Two-Oscillator Systems}

\author{Kevin K. Lin$^{1}$, Eric Shea-Brown$^{1,2}$, and Lai-Sang Young$^{1}$ \\ \\
$^1$ Courant Institute of Mathematical Sciences,  \\
$^2$ Center for Neural Science \\
New York University, New York, NY 10012, U.S.A. }

\begin{document}

\maketitle

\begin{abstract}
  This paper concerns the \textit{reliability} of a pair of coupled
  oscillators in response to fluctuating inputs.  Reliability means that
  an input elicits essentially identical responses upon repeated
  presentations regardless of the network's initial condition. Our main
  result is that both reliable and unreliable behaviors occur in this
  network for broad ranges of coupling strengths, even though individual
  oscillators are always reliable when uncoupled.  A new finding is that
  at low input amplitudes, the system is highly susceptible to unreliable
  responses when the feedforward and feedback couplings are roughly
  comparable. A geometric explanation based on   shear-induced chaos
  at the onset of phase-locking is proposed.
\end{abstract}

\begin{footnotesize}
\tableofcontents
\end{footnotesize}

\bigskip

\section*{Introduction}
\addcontentsline{toc}{section}{Introduction}

This paper, together with its companion paper {\it Reliability of
  Coupled Oscillators II}, contain a mathematical treatment of the
question of {\it reliability}.  Reliability here refers to whether a
system produces identical responses when it is repeatedly presented
with the same stimulus.  Such questions are relevant to signal
processing in biological and engineered systems.  Consider, for
example, a network of inter-connected neurons with some background
activity.  An external stimulus in the form of a time-dependent signal
is applied to this neural circuitry, which processes the signal and
produces a response in the form of voltage spikes.
We say the system is {\it reliable} if, independent of its
state at the time of presentation, the same stimulus elicits
essentially identical responses following an initial period of
adjustment.  That is, the response to a given signal is
reproducible~\cite{piko01,Bry+76,MS95,Nak+al05,Zho+03,Ter+03,Rit+03,Pak+01,Pak+03,Gol+05,Gol+06,Ter+07}.

This study is carried out in the context of (heterogeneous) networks of
interconnected oscillators.  We assume the input signal is received by
some components of the network and relayed to others, possibly in the
presence of feedback connections.  Our motivation
 is a desire to understand the connection between a
network's reliability properties and its architecture, {\em i.e.}  its
``circuit diagram,'' the strengths of various connections, {\em etc.}
This problem is quite different from the simpler and much studied
situation of uncoupled oscillators driven by a common input.  In the
latter, the concept of reliability coincides with synchronization, while
in coupled systems, internal synchronization among constituent
components is not a condition for reliability.

To simplify the analysis, we assume the constituent oscillators are {\it
  phase oscillators} or circle rotators, and that they are driven by a
fluctuating input which, for simplicity, we take to be white noise.
Under these conditions, systems consisting of a single, isolated phase
oscillator have been explored extensively and shown to be reliable;
see, {\it e.g.}, ~\cite{Ter+03,Rit+03}.  Our results are presented in a
two-part series:

\medskip
{Paper I. \ Two-oscillator systems

\medskip
Paper II. \ Larger networks}

\medskip
\noindent The present paper contains an in-depth analysis of a
2-oscillator system in which the stimulus is received by one of the
oscillators. Our results show that such a
system can support both reliable and unreliable dynamics depending on
coupling constants; a more detailed discussion of the main points of
this paper is given later in the Introduction.  Paper II extends some
of the ideas developed in this paper to larger networks that can be
decomposed into subsystems or ``modules'' so that the inter-module
connections form an acyclic graph.  At the level of inter-module
connections, such ``acyclic quotient networks'' have essentially
feedforward dynamics; they are 
 reliable if all the component modules
are also reliable.  Acyclic quotient networks are allowed to contain
unreliable modules, however, and the simplest example of such a module
is, as shown in this paper, an oscillator pair with nontrivial
feedforward-feedback connections.  In Paper II, we also explore issues
surrounding the creation and propagation of unreliability in larger
networks.

With regard to aims and methodology, we seek to identify and explain
relevant phenomena, and to make contact with rigorous mathematics
whenever we can, hoping in the long run to help bring dynamical systems
theory closer to biologically relevant systems.  The notion of
reliability, for example, is prevalent  in neuroscience.
With all the idealizations and
simplifications at this stage of our modeling, we do not expect our
results to be directly applicable to real-life situations, but hope
that on the phenomenological level they will shed light on systems that
share some characteristics with the oscillator networks studied here.
Rigorous mathematical
results are presented in the form of ``Theorems,'' ``Propositions,''
etc., and simulations are used abundantly. Our main results are
qualitative. They are a combination of rigorous statements and
heuristic explanations supported by numerical simulations and/or
theoretical understanding.

\bigskip
\noindent {\bf Content of present paper}

\medskip
\noindent A motivation for this work is the following naive (and partly
rhetorical) question: {\em Are networks of coupled phase oscillators
  reliable, and if not, how large must a network be to exhibit
  unreliable behavior?}  Our answer to this question is that unreliable
behavior occurs already in the 2-oscillator configuration in Diagram
(\ref{two-cell schematic}).  Our results demonstrate clearly that such a
system can be reliable or unreliable, and that both types of behaviors
are quite prominent, depending in a generally predictable way on the
nature of the feedforward and feedback connections. Furthermore, we
identify geometric mechanisms responsible for these behaviors.

Referring the reader again to Diagram (\ref{two-cell schematic}), three
of the system's parameters are $\aff$ and $\afb$, the feedforward and
feedback coupling constants, and $\varepsilon$, the amplitude of the
stimulus.  The following is a preview of the main points of this paper:

\begin{enumerate}
\item {\em Lyapunov exponents as a measure of reliability.}  Viewing the
  stimulus-driven  system as described by a stochastic differential equation and
  leveraging existing results from random dynamical systems theory, we
  explain in Sect.~2 why the top Lyapunov exponent ($\lmax$) of the
  associated stochastic flow is a reasonable measure of reliability. In
  this interpretation, reliability is equated with $\lmax<0$, which is
  known to be equivalent to having {\it random sinks} in the dynamics,
  while unreliability is equated with $\lmax>0$, which is equivalent to
  the presence of {\it random strange attractors}.

\item {\em Geometry and zero-input dynamics.} In pure feedforward and
  feedback configurations, {\it i.e.}, at $\afb=0$ or $\aff=0$,
we identify geometric structures that prohibit unreliability.
  Our main result about zero-input systems is on {\em
    phase-locking}: in Sect. 3.3, we prove that for all
  $\aff$ in a broad range,
   the system is 1:1 phase-locked for either $\afb \gtrsim \aff$
  or $\afb \lesssim \aff$ depending on the relative intrinsic
  frequencies of the two oscillators.  This phenomenon has important
  consequences for reliability.


\item {\em Shear-induced chaos as main cause for unreliability at low
    drive amplitudes.} Recent advances in dynamical systems theory have
  identified a mechanism for producing chaos via the interaction of a
  forcing with the underlying shear in a system.  The dynamical
  environment near the onset of phase-locking is particularly
  susceptible to this mechanism. We verify in Sect. 4.2 that the
  required ``shearing'' is indeed present in our 2-oscillator system.
  Applying the cited theory, we are able to predict the reliability or
  lack thereof for coupling parameters near the onset of phase-locking.
  At low drive amplitudes, this is the primary cause for unreliability.

\item {\em Reliability profile as a function of $\aff, \afb$ and
    $\varepsilon$.} Via numerical simulations and theoretical reasoning,
  we deduce a rough reliability profile for the 2-oscillator system as a
  function of the three parameters above. With the increase in $|\aff|,
  |\afb|$ and/or $\varepsilon$, both reliable and unreliable regions
  grow in size and become more robust, meaning $\lmax$ is farther away
  from zero. The main findings are summarized in Sect. 4.3.

\end{enumerate}


\section{Model and Formulation}

\subsection{Description of the model}
\label{Description of the model}

We consider in this paper a small network consisting of two coupled
phase oscillators forced by an external stimulus as shown:
\begin{equation}
\setlength{\unitlength}{0.0004in}
\begingroup\makeatletter\ifx\SetFigFont\undefined%
\gdef\SetFigFont#1#2#3#4#5{%
  \reset@font\fontsize{#1}{#2pt}%
  \fontfamily{#3}\fontseries{#4}\fontshape{#5}%
  \selectfont}%
\fi\endgroup%
{\renewcommand{\dashlinestretch}{30}
\begin{picture}(4533,1113)(0,-10)
\put(2625,849){\makebox(0,0)[lb]{{\SetFigFont{10}{14.4}{\rmdefault}{\mddefault}{\updefault}$\aff$}}}
\put(3916.875,502.181){\arc{808.949}{5.2529}{7.4711}}
\blacken\path(4236.560,795.575)(4125.000,849.000)(4198.294,749.361)(4236.560,795.575)
\put(1575,549){\ellipse{1082}{1082}}
\put(3900,549){\ellipse{1082}{1082}} \thicklines
\path(2025,549)(2175,549) \path(4350,549)(4500,549) \thinlines
\path(3225,399)(2325,399)
\blacken\path(2445.000,429.000)(2325.000,399.000)(2445.000,369.000)(2445.000,429.000)
\path(2325,699)(3225,699)
\blacken\path(3105.000,669.000)(3225.000,699.000)(3105.000,729.000)(3105.000,669.000)
\put(3825,474){\makebox(0,0)[lb]{{\SetFigFont{10}{14.4}{\rmdefault}{\mddefault}{\updefault}$\theta_2$}}}
\put(1500,474){\makebox(0,0)[lb]{{\SetFigFont{10}{14.4}{\rmdefault}{\mddefault}{\updefault}$\theta_1$}}}
\put(-600,450){\makebox(0,0)[lb]{{\SetFigFont{10}{14.4}{\rmdefault}{\mddefault}{\updefault}$\eps~I(t)$}}}
\put(200,400){\makebox(0,0)[lb]{{\SetFigFont{10}{14.4}{\rmdefault}{\mddefault}{\updefault}{\huge
    $\rightsquigarrow$}}}}
\put(2625,99){\makebox(0,0)[lb]{{\SetFigFont{10}{14.4}{\rmdefault}{\mddefault}{\updefault}$\afb$}}}
\put(1573.875,549.181){\arc{808.949}{5.2529}{7.4711}}
\blacken\path(1893.560,842.575)(1782.000,896.000)(1855.294,796.361)(1893.560,842.575)
\end{picture}
}

\label{two-cell schematic}
\end{equation}
We assume the coupling is via smooth pulsatile interactions as in
\cite{TH98,geomtime,erme96,gerstner96what}, and the equations defining
the system are given by
\begin{equation}
\begin{array}{rcl}
  \dot\theta_1 &=& \omega_1 + \afb\ z(\theta_1)\ g(\theta_2)
  + \varepsilon z(\theta_1)\ I(t)\ ,\\
  \dot\theta_2 &=& \omega_2 + \aff\ z(\theta_2)\
  g(\theta_1)\ .\\
\end{array}
\label{2cell}
\end{equation}
The state of each oscillator is described by a phase, {\em i.e.}, an
angular variable $\theta_i \in {\mathbb S}^1 \equiv {\mathbb R}/{\mathbb
  Z}, i=1,2$.  The constants $\omega_1$ and $\omega_2$ are the cells'
intrinsic frequencies.  We allow these frequencies to vary, but assume
for definiteness that they are all within about $10\%$ of $1$.  The
external stimulus is denoted by $\eps I(t)$. Here $I(t)$ is taken to be
white noise, and $\eps$ is the signal's amplitude. Notice that the
signal is received only by cell 1.  The coupling is via a ``bump
function'' $g$ which vanishes outside of a small interval $(-b,b)$.  On
$(-b,b)$, $g$ is smooth, it is $\geq 0$, has a single peak, and
satisfies $\int_{-b}^b{g(\theta)\ d\theta} = 1$. The meaning of $g$ is
as follows: We say the $i$th oscillator ``spikes'' when
$\theta_i(t)=0$. Around the time that an oscillator spikes, it emits a
pulse which modifies the other oscillator.  The sign and magnitude of
the feedforward coupling is given by $\aff$ (``forward'' refers to the
direction of stimulus propagation): $\aff>0$ (resp. $\aff <0$) means
oscillator 1 excites (resp. inhibits) oscillator 2 when it spikes.  The
feedback coupling, $\afb$, plays the complementary role. In this paper,
$b$ is taken to be about $\frac{1}{20}$, and $\aff$ and $\afb$ are taken
to be ${\mathcal O}(1)$.  Finally, $z(\theta)$, often called the {\em
  phase response curve}~\cite{Winf74,G75,erme96,BHMphase,Krev}, measures
the variable sensitivity of an oscillator to coupling and stimulus
input. In this paper, we take $z(\theta)
=\frac{1}{2\pi}(1-\cos(2\pi\theta))$ (see below).

\subsection{Neuroscience interpretations}

Coupled phase oscillators arise in many settings
\cite{piko01,SS00,Krev,HopIzhik,geomtime,excit}.
 Here, we briefly discuss their use in mathematical
neuroscience.

We think of phase oscillators as paradigms for systems with rhythmic
behavior.  Such models are often derived as limiting cases of
oscillator models in two or more dimensions. In particular, the
specific form of $z(\cdot)$ chosen here corresponds to the normal form
for oscillators near saddle-node bifurcations on their limit cycles
\cite{erme96}. This situation is typical in neuroscience, where neural
models with $z(\tht) \approx 1-\cos(\cdot)$ are referred to as ``Type
I.'' The pulse- or spike-based coupling implemented by $g(\cdot)$ may
also be motivated by the synaptic impulses sent between neurons after
they fire action potentials (although this is not the only setting in
which pulsatile coupling arises
\cite{geomtime,BHMsiro,EKmult,TH98,StrMir90,HopHer95,NumPer85,peskin,HopIzhik}).

The general conclusions that we will present do not depend on the
specific choices of $z(\cdot)$ and $g(\cdot)$, but rather on their
qualitative features.  Specifically, we have checked that our main
results about reliability and phase locking are essentially unchanged
when the $z(\cdot)$ function becomes asymmetric and the location of the
$g(\cdot)$ impulse is somewhat shifted, as would correspond more
closely to neuroscience \cite{erme96}.  Therefore, the behavior we find
here can be expected to be fairly prototypical for pairs of
pulse-coupled Type I oscillators.

\begin{figure}
\begin{center}
\resizebox{4.5in}{!}{\includegraphics[bb=0in 0in 9.6in 3in]{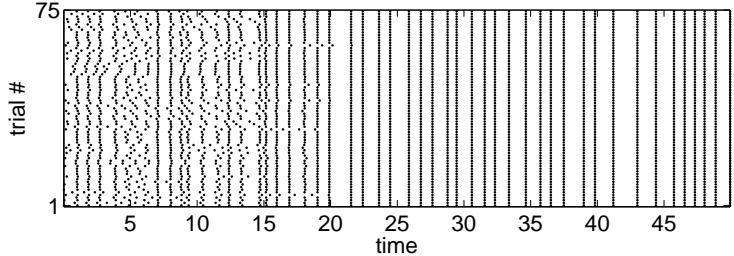}}
\end{center}
\caption{Raster plot for an isolated oscillator.  In each experiment, 75
  trials are performed, and dots are placed at spike times.  Nearly
  identical spike times are observed after a transient, indicating
  reliability.}
\label{f.raster1}
\end{figure}

A standard way to investigate the reliability of a system of spiking
oscillators --- both in the laboratory and in simulations --- is to
repeat the experiment using a different initial condition each time but
driving the system with {\it the same} stimulus $\eps I(t)$, and to
record spike times in raster plots.  Fig.~\ref{f.raster1} shows such a
raster plot for an isolated oscillator of the present type, as studied
by~\cite{Rit+03,Gut+03}.  Note that, for each repetition, the oscillator produces
essentially identical spike times after a transient, demonstrating its
reliability.



\section{Measuring Reliability}

All of the ideas discussed in this section are general and
are easily adapted to larger networks.

\subsection{A working definition of reliability} \label{s.wkdef}

We attempt to give a formal definition of reliability in a general
setting. Consider a dynamical system on a domain $M$ (such as a manifold
or a subset of Euclidean space).  A signal in the form of
$\iota(t) \in {\mathbb R}^n, t \in [0,\infty),$ is presented to the system.
The response $F(t)$ of the
system to this signal is given by $F(t) = F(x_0, t, \{\iota(s)\}_{0 \leq
  s < t})$.  That is to say, the response at time $t$ may, in principle,
depend on $x_0 \in M$, the initial state of the system when the signal
is presented, and the values of the signal up to time $t$.

In the model as described in Sect. 1.1, $F(t)$ can be thought of as the pair
$(\theta_1(t), \theta_2(t))$ or $\Psi(\theta_1(t),\theta_2(t))$, the
value of an observable at time $t$.

We propose now one way to define reliability.  Given a dynamical system,
a class of signals $\cal I$, and a response function $F$, we say the
system is {\it reliable} if for almost all $\iota \in {\cal I}$ and
$x_0, x_0' \in M$,
$$\|F(x_0, t, \{\iota(s)\}_{0 \leq s < t})-F(x_0', t, \{\iota(s)\}_{ 0 \leq s < t})\|
\to 0 \quad {\rm as }\quad t \to \infty\ .
$$
Here $\| \cdot\|$ is a norm on the range space of $F$.  We do not
claim that this is the only way to capture the idea of reliability,
but will use it as our operational definition.

We point out some of the pitfalls of this definition: In practice,
signals are never presented for infinite times, and in some situations,
responses can be regarded as reliable only if the convergence above is
rapid.  By the same token, not all initial conditions are equally
likely, leaving room for probabilistic interpretations.

Finally, one should not expect unreliable responses to be
fully random. On the contrary, as we will show in Sect. 2.2,
they tend to possess a great deal of structure, forming what
are known as {\it random strange attractors}.


\subsection{Reliability, Lyapunov exponents, and random attractors}

We discuss here some mathematical tools that can be used to quantify how
reliable or unreliable a driven system is.  With $I(t)$ taken to be
realizations of white noise, (\ref{2cell}) can be put into the framework
of a {\it random dynamical system} (RDS).  We begin by reviewing some
relevant mathematics~\cite{RDS,Bax92}.  Consider a general stochastic
differential equation
\begin{equation}
\label{eq:sde} dx_t = a(x_t)\ dt + \sum_{i=1}^k b_i(x_t)\circ dW^i_t\ .
\end{equation}
In this general setting, $x_t \in M$ where $M$
is a compact Riemannian manifold,
the $W^i_t$ are independent standard Brownian motions, and
the equation is of Stratonovich type. Clearly, Eq.~(\ref{2cell})
is a special case of Eq.~(\ref{eq:sde}): $x_t$ describes the phases of
the $2$ oscillators,
$M={\mathbb T}^2 \equiv {\mathbb{S}}^1 \times {\mathbb{S}}^1$ (thus
we have the choice between the It\^{o} or Stratonovich integral),
and $k=1$.

In general, one fixes an initial $x_0$, and looks at the distribution
of $x_t$ for $t>0$. Under fairly general conditions, these
distributions converge to a unique stationary measure $\mu$, the
density of which is given by the Fokker-Planck equation. Since
reliability is about a system's reaction to a single stimulus, {\it i.e.},
a single realization of the $W^i_t$, at a time, and concerns
the simultaneous evolution of all or large ensembles
of initial conditions, of relevance to us are not the
distributions of $x_t$ but {\it flow-maps} of
the form $F_{t_1,t_2;\omega}$.  Here $t_1<t_2$ are two points in time,
$\omega$ is a sample Brownian path, and
$F_{t_1,t_2;\omega}(x_{t_1})=x_{t_2}$ where $x_t$ is the solution of
(\ref{eq:sde}) corresponding to $\omega$.  A well known theorem states
that such {\it stochastic flows of diffeomorphisms} are well defined if
the functions $a(x)$ and $b(x)$ in Eq. (\ref{eq:sde}) are sufficiently
smooth. More precisely, the maps $F_{t_1,t_2;\omega}$ are well defined
for almost every $\omega$, and they are invertible, smooth
transformations with smooth inverses.  Moreover, $F_{t_1,t_2;\omega}$
and $F_{t_3,t_4;\omega}$ are independent for $t_1<t_2<t_3<t_4$. These
results allow us to treat the evolution of systems described by
(\ref{eq:sde}) as compositions of random, {\it i.i.d.}, smooth maps.
Many of the techniques for analyzing smooth deterministic systems have
been extended to this random setting. We will refer to the resulting
body of work as RDS theory.

Similarly, the stationary measure $\mu$, which gives  the
steady-state distribution
averaged over all realizations $\omega$, does not describe what we see
when studying a system's reliability. Of relevance are the {\it sample
measures} $\{\mu_\omega\}$, which are the conditional measures of $\mu$
given the past. Here we think of $\omega$ as defined for all
$t \in (-\infty,\infty)$ and not just for $t>0$. Then $\mu_\omega$ describes what one sees at
$t=0$ given that the system has experienced the input defined by
$\omega$ for all $t <0$. This is an idealization: in reality, the input
is presented on a time interval $[-t_0, 0)$ for some $t_0>0$. Two
useful facts about these sample measures are
\smallskip
\begin{itemize}
\item[(a)]
$(F_{-t,0;\omega})_*\mu \to \mu_\omega$ as $t \to \infty$, where
$(F_{-t,0;\omega})_*\mu$ is the measure obtained by transporting $\mu$
forward by $F_{-t,0;\omega}$, and

\smallskip
\item[(b)]
the family $\{\mu_\omega\}$ is invariant in the sense that
$(F_{0,t;\omega})_*(\mu_\omega) = \mu_{\sigma_t(\omega)}$ where
$\sigma_t (\omega)$ is the time-shift of the sample path $\omega$ by
$t$.
\end{itemize}
\smallskip
\noindent Thus, if our initial distribution is given by a probability density
$\rho$ and we apply the stimulus corresponding to $\omega$, then the
distribution at time $t$ is $(F_{0,t;\omega})_*\rho$. For $t$
sufficiently large, one expects in most situations that
$(F_{0,t;\omega})_*\rho$ is very close to $(F_{0,t;\omega})_*\mu$,
which by (a) above is essentially given by $\mu_{\sigma_t(\omega)}$.
The time-shift by $t$ of $\omega$ is necessary because by definition,
$\mu_\omega$ is the conditional distribution of $\mu$ at time $0$.

\begin{figure}[t]
\begin{center}
\begin{tabular}{cccc}
$t=20$&$t=50$&$t=500$&$t=1900$\\
\resizebox{1.7in}{!}{\includegraphics*[bb=0 0 590 590]{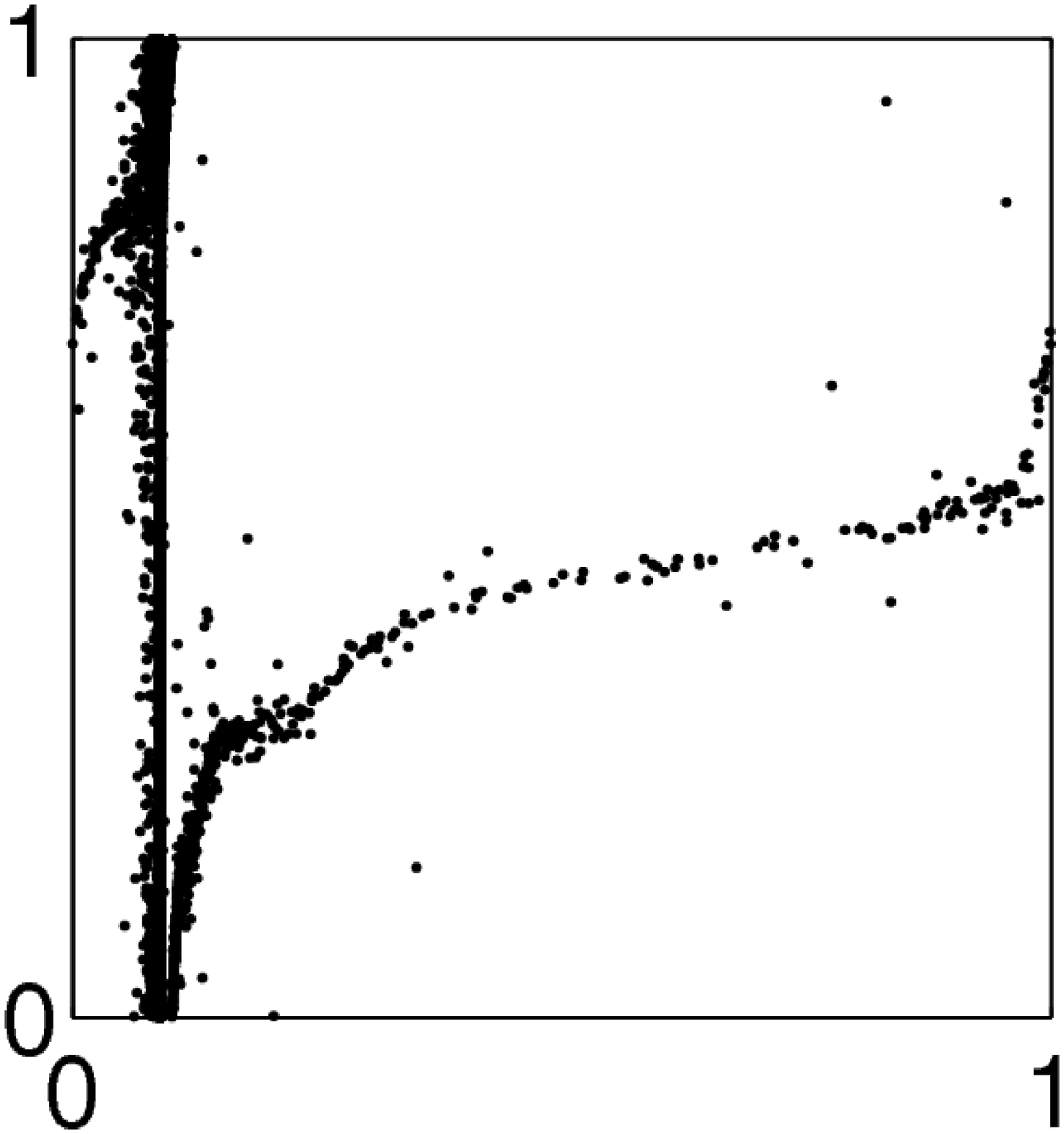}}\hspace*{-12pt}&
\resizebox{1.7in}{!}{\includegraphics*[bb=0 0 590 590]{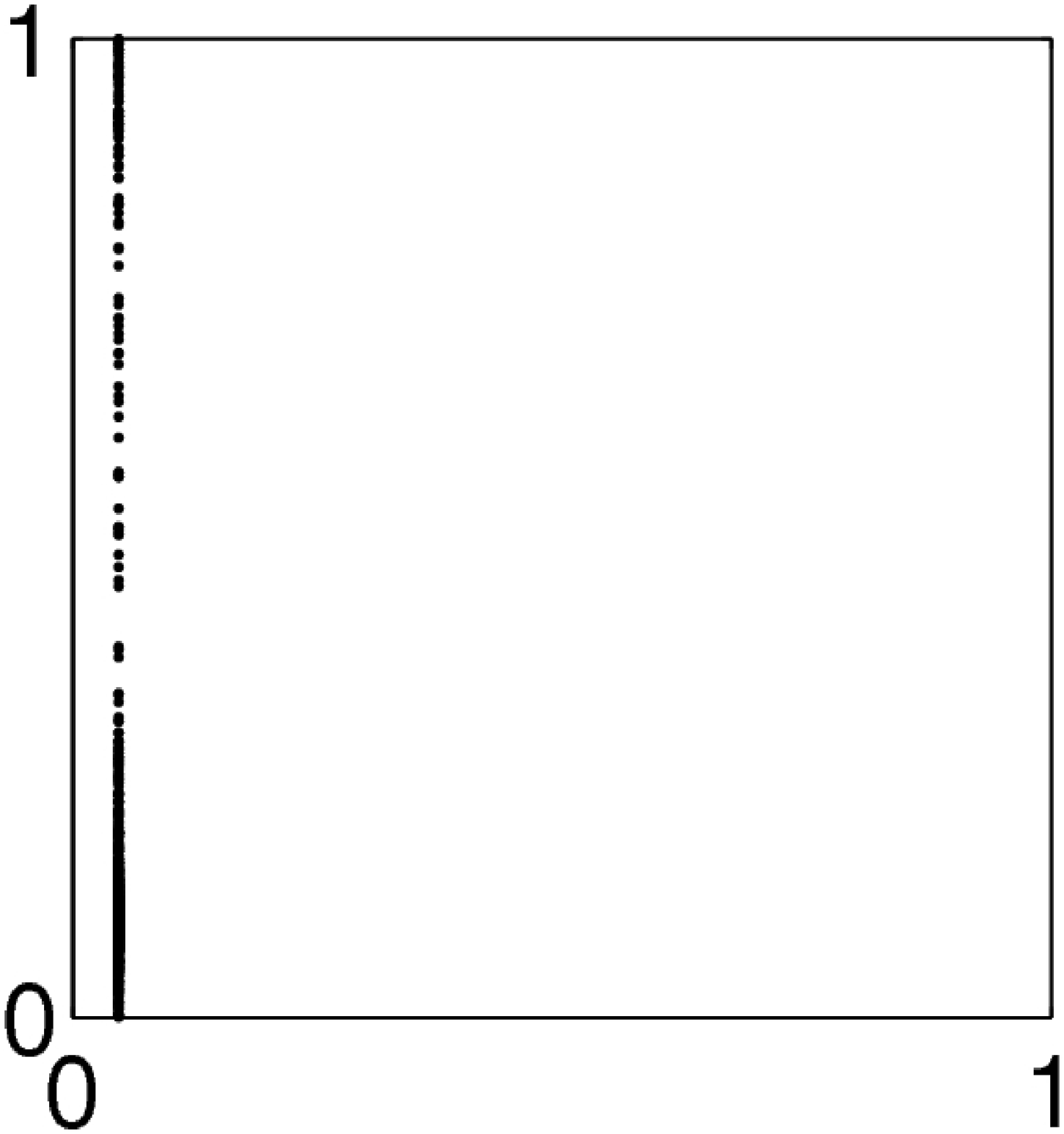}}\hspace*{-14pt}&
\hspace*{9pt}\resizebox{1.675in}{!}{\includegraphics*[bb=0 0 215 231]{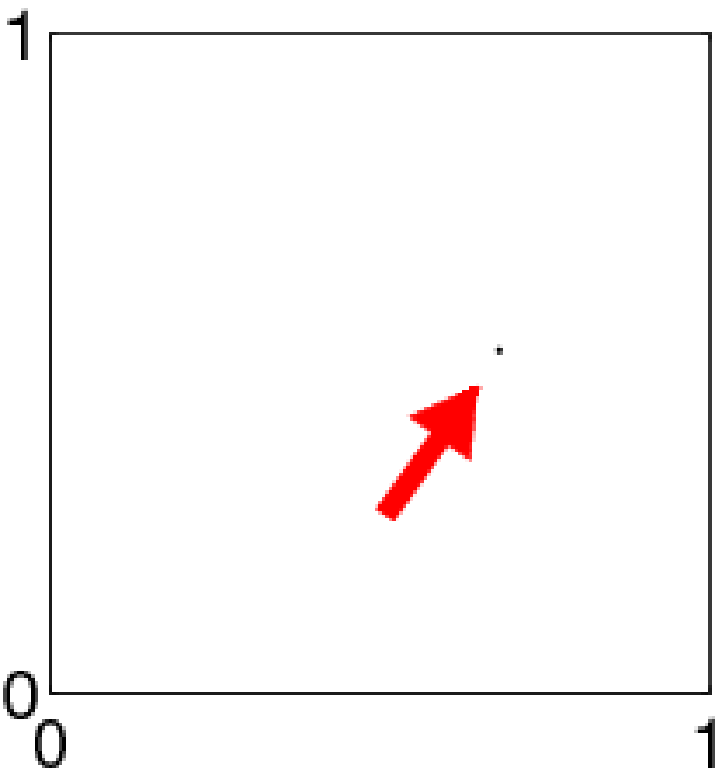}}\hspace*{-14pt}&
\resizebox{1.675in}{!}{\includegraphics*[bb=0 0 215 231]{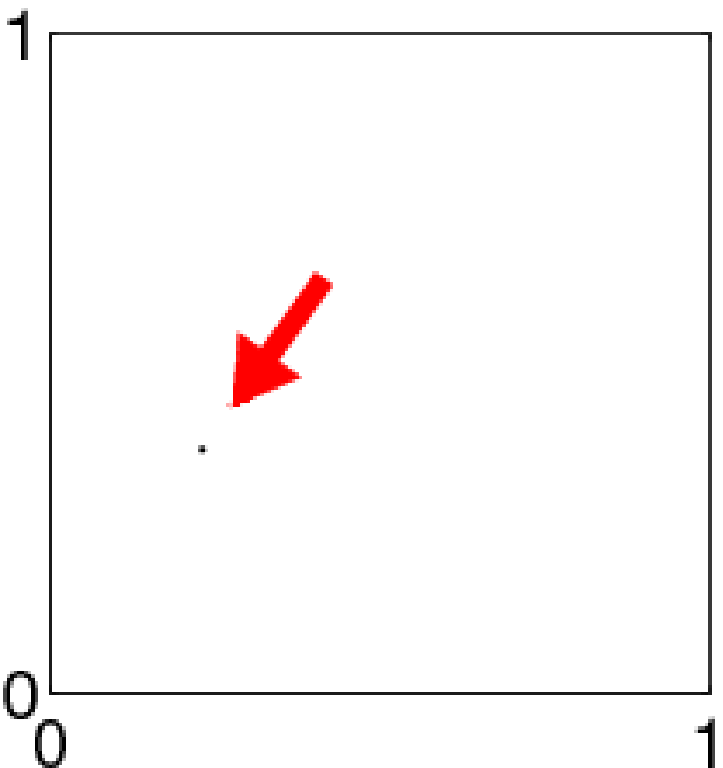}}\\
\end{tabular}\\ \vspace{.1in}
(a) Random fixed point ($\lmax < 0$)\\\vspace{12pt}
\begin{tabular}{cccc}
$t=20$&$t=50$&$t=500$&$t=1900$\\
\resizebox{1.7in}{!}{\includegraphics*[bb=0 0 590 590]{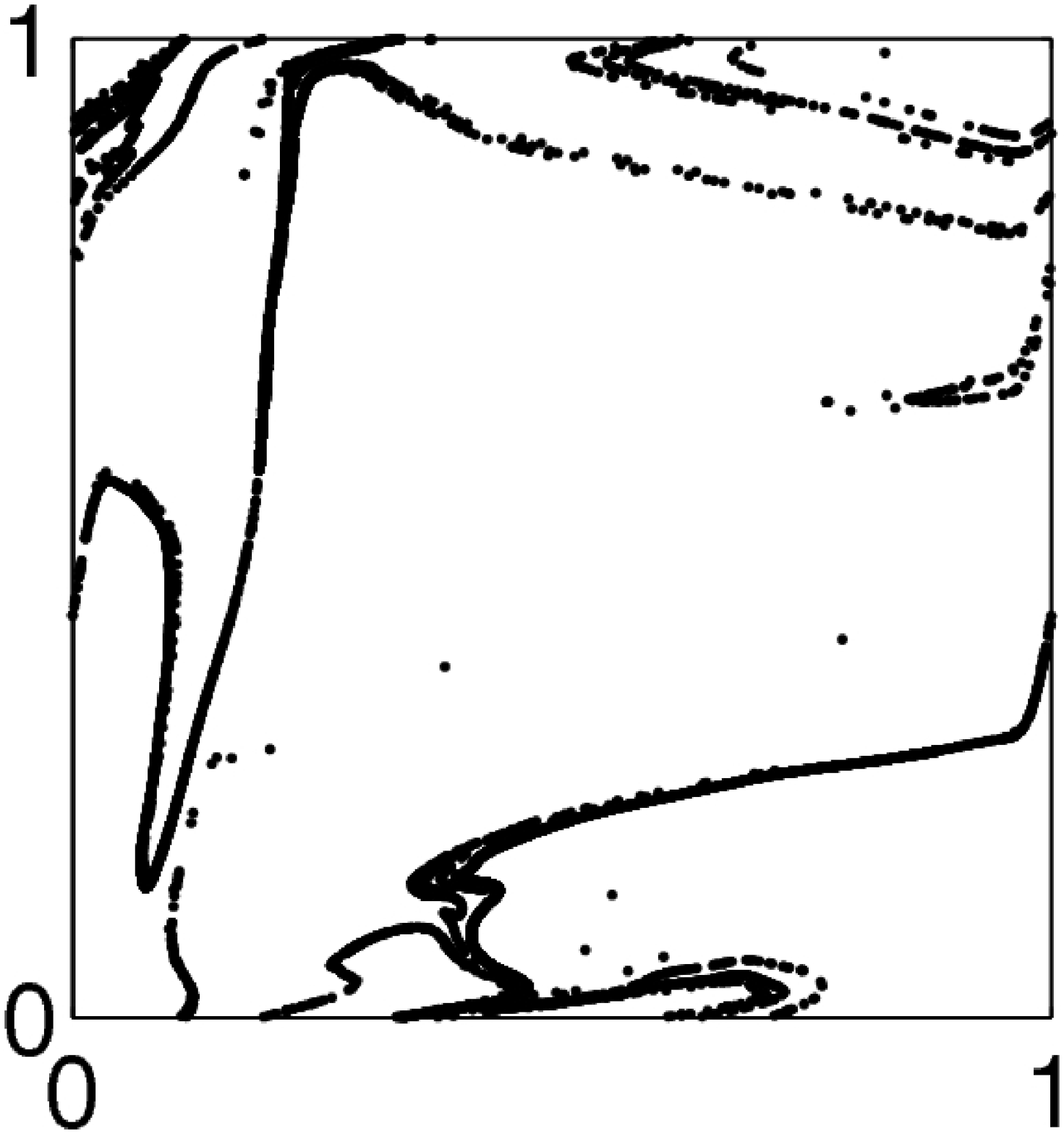}}\hspace*{-14pt}&
\resizebox{1.7in}{!}{\includegraphics*[bb=0 0 590 590]{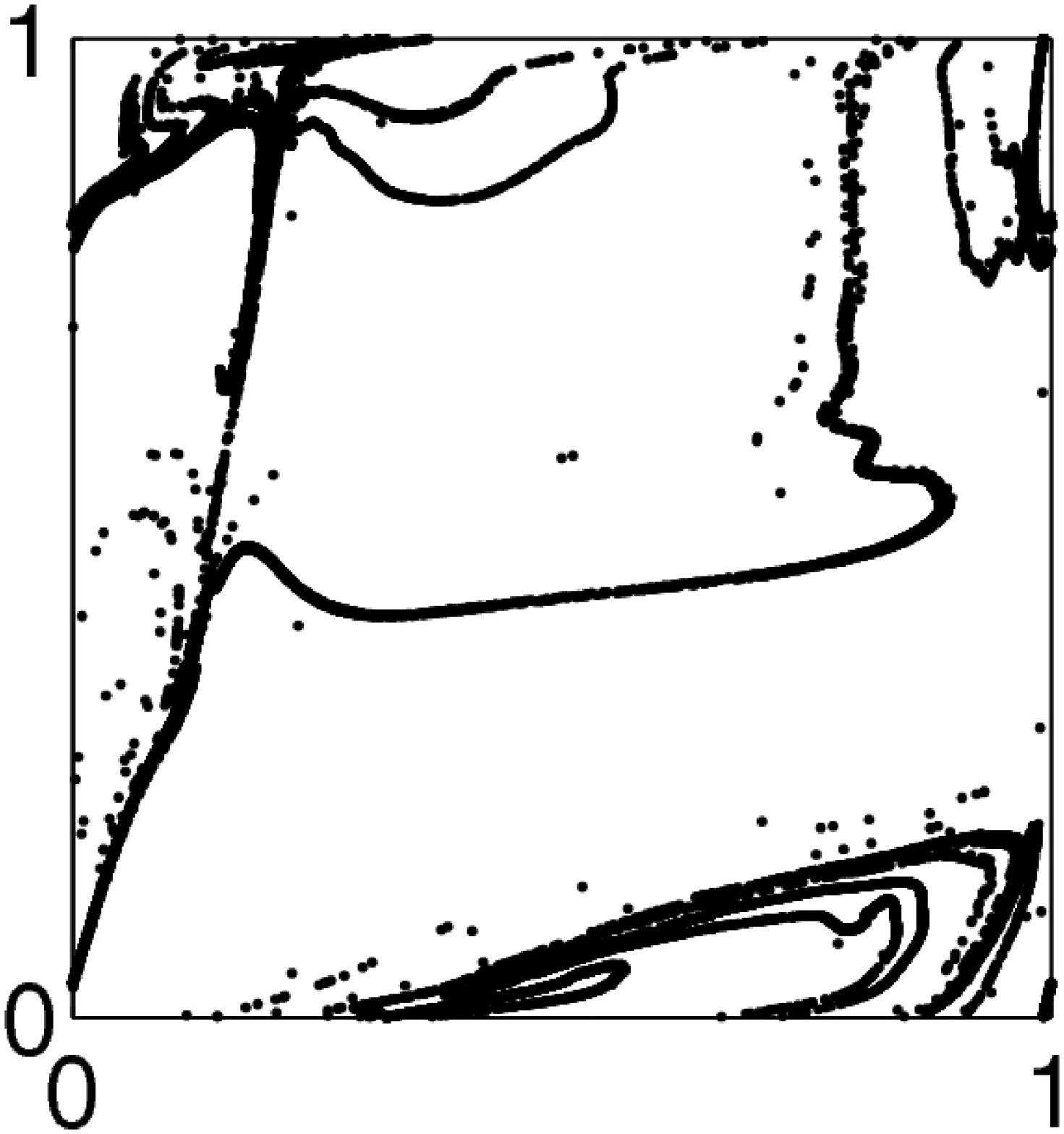}}\hspace*{-14pt}&
\resizebox{1.7in}{!}{\includegraphics*[bb=0 0 590 590]{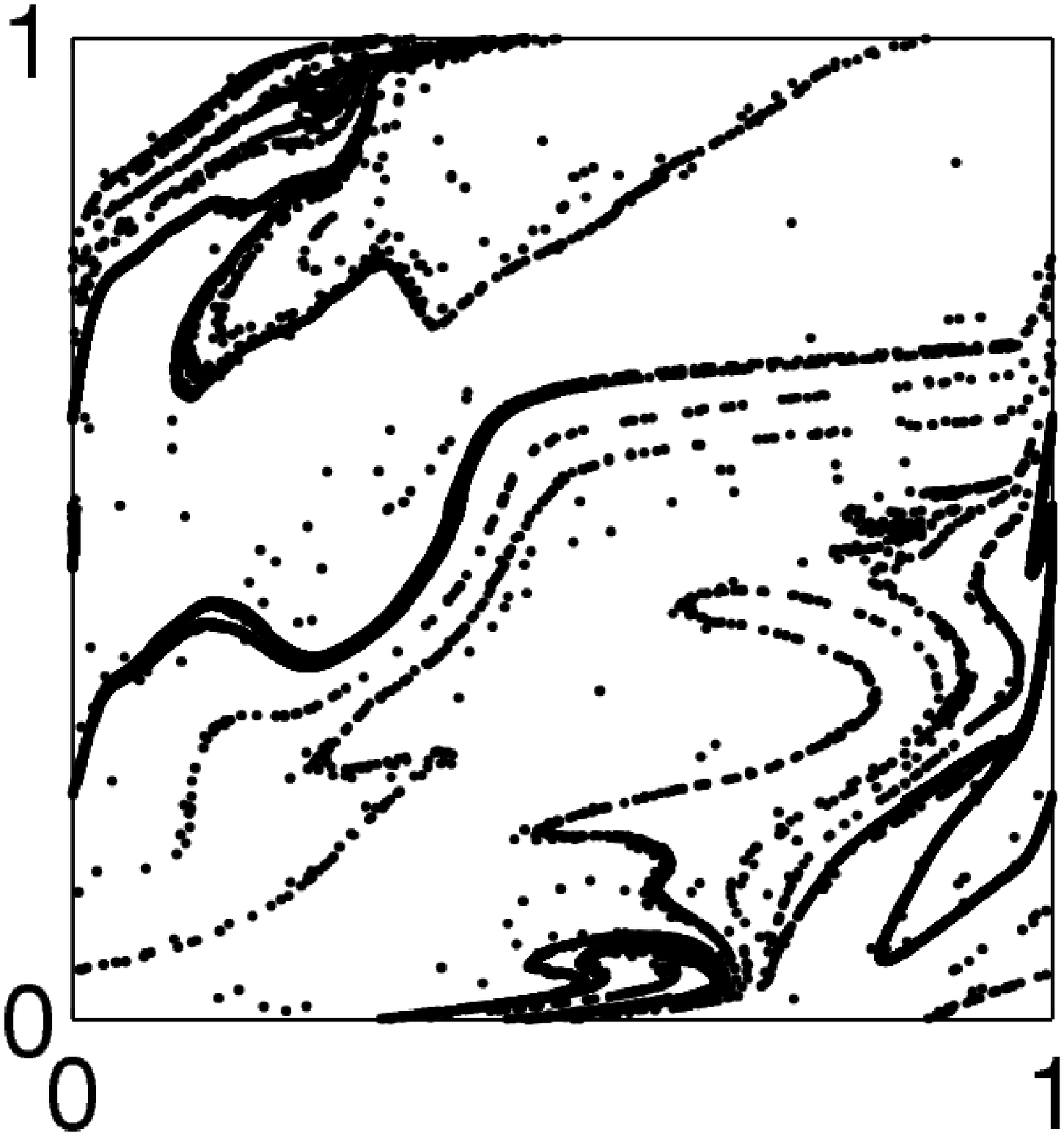}}\hspace*{-14pt}&
\resizebox{1.7in}{!}{\includegraphics*[bb=0 0 580 600]{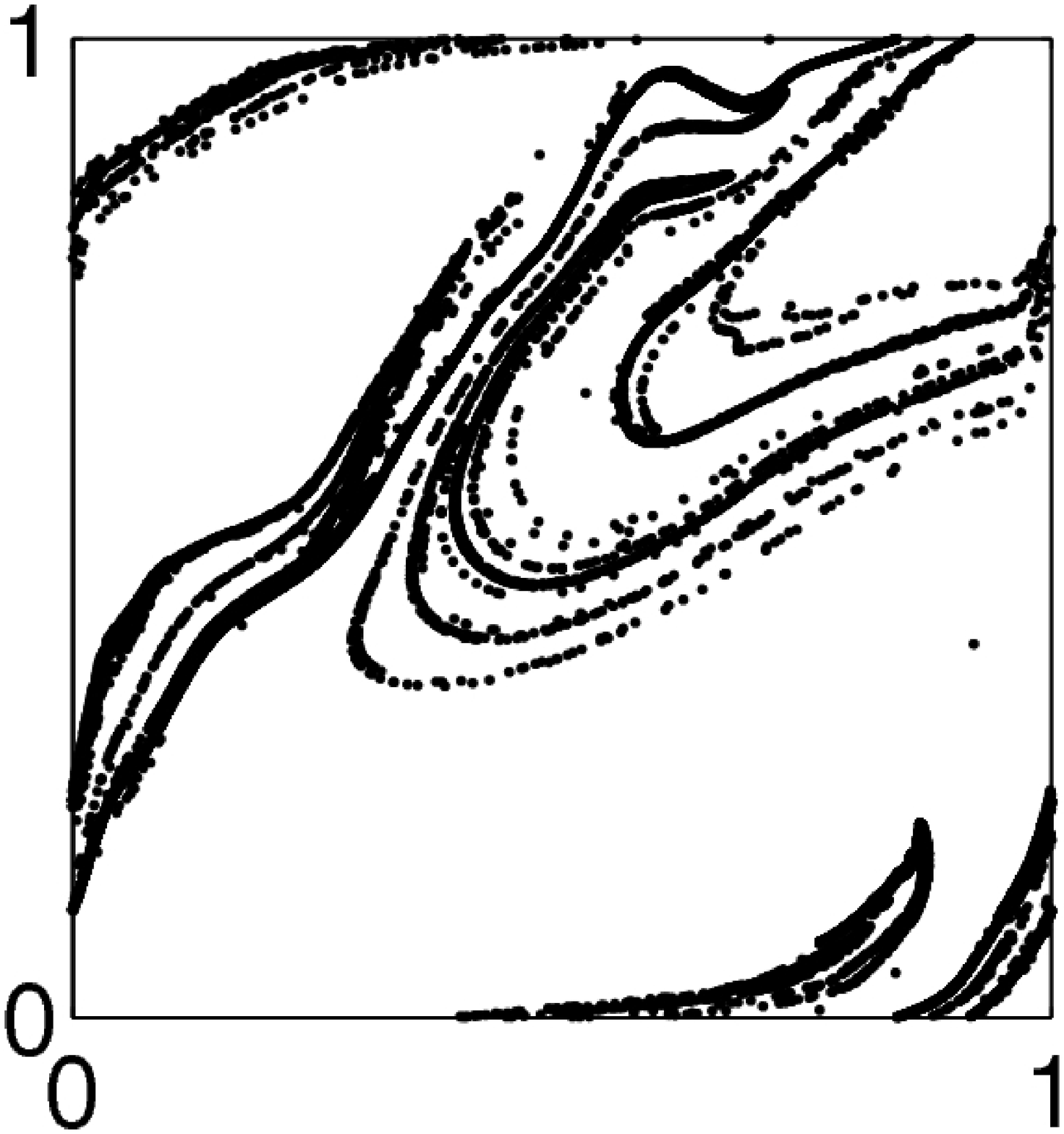}}\\
\end{tabular}\\
(b) Random strange attractor ($\lmax > 0$)
\end{center}
\caption{Snapshots of sample measures for Eq.~(\ref{2cell})
at various times in response to a single realization of the stimulus.
Two different sets of parameters are used in (a) and (b).
In (a), the sample measures converge
to a random fixed point.  In (b), the sample measures
  converge to a random strange attractor.  See Theorem 1.}
\label{f.snapshot}
\end{figure}

Fig.~\ref{f.snapshot} shows some snapshots of $(F_{0,t;\omega})_*\rho$
for the coupled oscillator pair in the system (1) for two different sets
of parameters.  In the panels corresponding to $t\gg 1$, the
distributions approximate $\mu_{\sigma_t(\omega)}$.
In these simulations, the
initial distribution $\rho$ is the stationary density of
Eq.~(\ref{2cell}) with $\eps\approx 0.01$, the interpretation being that
the system is intrinsically noisy even in the absence of external
stimuli. Observe that these pictures evolve with time, and as $t$
increases they acquire similar qualitative properties depending on the
underlying system. This is in agreement with RDS theory, which tells us
in fact that the $\mu_{\sigma_t(\omega)}$ obey a statistical law for
almost all $\omega$.  Observe also the strikingly different behaviors in
the top and bottom panels. We will follow up on this observation
presently.  First, we recall two mathematical results that describe the
dichotomy.

In deterministic dynamics, Lyapunov exponents measure the exponential
rates of separation of nearby trajectories. Let $\lambda_{\max}(x)$ denote
the largest Lyapunov exponent along the trajectory starting from $x$. Then a positive
$\lambda_{\max}$ for a large set of initial conditions is generally thought
of as synonymous with
chaos, while the presence of stable equilibria is characterized by
$\lambda_{\max}<0$. For smooth random dynamical systems, Lyapunov
exponents are also known to be well defined; moreover, they are {\it
nonrandom}, {\em i.e.} they do not depend on $\omega$. If, in addition, the
invariant measure is erogdic, then $\lambda_{\max}$ is constant
almost everywhere in the phase space.
In Theorem 1 below, we present two results from RDS
theory that together suggest that the sign of $\lmax$ is a good
criterion for distinguishing between reliable and unreliable behavior:

\begin{theorem} In the setting of Eq.~(\ref{eq:sde}), let
$\mu$ be an ergodic stationary measure.
\begin{itemize}
\item[(1)]{\bf (Random sinks)} {\rm \cite{LeJ85}} If
  $\lambda_{\max}<0$, then with probability 1, $\mu_\omega$ is supported
  on a finite set of points.
\item[(2)]{\bf (Random strange attractors)} {\rm \cite{Led+88}}
If $\mu$ has a density and $\lambda_{\max}>0$, then with probability 1,
$\mu_\omega$ is a random SRB measure.
\end{itemize}
\end{theorem}

The conclusion of Part (1) is interpreted as follows: The scenario in
which the support of $\mu_\omega$ consists of a single point corresponds
exactly to reliability for almost every $\omega$ as defined in
Sect.~\ref{s.wkdef}.  This was noted in, e.g., \cite{Rit+03,Pak+01}.  For the $2$-oscillator system defined by
Eq.~(\ref{2cell}), the collapse of all initial conditions to a point is
illustrated in Fig.~\ref{f.snapshot}(a); notice that
$\mu_{\sigma_t\omega}$ continues to evolve as $t$ increases.  Even
though in general $\mu_\omega$ can be supported on more than one point
when $\lmax<0$, this seems seldom to be the case except in the presence
of symmetries.  We do not know of a mathematical result to support this
empirical observation, however.\footnote{Analysis of single-neuron
  recordings have revealed firing patterns which suggest the possible
  presence of random sinks with $> 1$ point \cite{Fel+04}.}  In the rest
of this paper, we will take the liberty to equate $\lmax<0$ with
reliability.

The conclusion of Part (2) requires clarification: In deterministic
dynamical systems theory, SRB measures are natural invariant measures
that describe the asymptotic dynamics of chaotic dissipative systems (in
the same way that Liouville measures are the natural invariant measures
for Hamiltonian systems). SRB measures are typically singular.  They are
concentrated on unstable manifolds, which are families of curves,
surfaces etc. that wind around in a complicated way in the phase space
{\cite{Eck+85}}. Part (2) of Theorem 1 generalizes these ideas to random
dynamical systems. Here, random (meaning
  $\omega$-dependent) SRB measures live on random unstable
manifolds, which are complicated families of curves, surfaces, {\em
  etc.}  that evolve with time. In particular, in a system with random
SRB measures, different initial conditions lead to very different
outcomes at time $t$ when acted on by the same stimulus; this is true
for all $t>0$, however large. It is, therefore, natural to regard
$\lambda_{\max}>0$ and the distinctive geometry of random SRB measures
as a signature of unreliability.

In the special case where the phase space is a circle, such as in the
case of a single oscillator, that the Lyapunov exponent $\lambda$ is
$\leq 0$ is an immediate consequence of Jensen's Inequality. In more
detail,
$$
\lambda(x) \ = \lim_{t \rightarrow \infty} \frac{1}{t} \log
F'_{0,t;\omega}(x)
$$
for typical $\omega$ by definition.  Integrating over initial conditions $x$,
we obtain
$$
\lambda \ = \int_{{\mathbb S}^1}\lim_{t \rightarrow \infty}
\frac{1}{t} \log F'_{0,t;\omega}(x)\ dx = \lim_{t \rightarrow \infty}
\frac{1}{t} \int_{{\mathbb S}^1} \log F'_{0,t;\omega}(x)\ dx \ .
$$
The exchange of integral and limit is permissible because
the required integrability conditions
are satisfied in stochastic flows~\cite{kifer}. Jensen's Inequality then
gives
\begin{equation}
\int_{{\mathbb S}^1} \log F'_{0,t;\omega}(x)\ dx \ \le \ \log
\int_{{\mathbb S}^1} F'_{0,t;\omega}(x)\ dx \ = \ 0\ .
\label{lyap}
\end{equation}
The equality above follows from the fact that $F_{0,t;\omega}$ is a
circle diffeomorphism. Since the gap in the inequality in (\ref{lyap})
is larger when $F'_{0,t;\omega}$ is farther from being a constant
function, we see that $\lambda<0$ corresponds to $F'_{0,t;\omega}$
becoming ``exponentially uneven'' as $t \to \infty$.  This is consistent
with the formation of random sinks.

The following results from general RDS theory shed light on the
situation when the system is multi-dimensional:

\begin{proposition} {\rm (see {\it e.g.} {\rm \cite{kifer}})} In the setting of
Eq.~(\ref{eq:sde}), assume $\mu$ has a density, and let
$\{\lambda_1,  \cdots, \lambda_d\}$ be the Lyapunov
exponents of the system counted with multiplicity. Then

\smallskip
(i) $\sum_i \lambda_i \leq 0$;

(ii) $\sum_i \lambda_i=0$ if and only if $F_{s,t,\omega}$ preserves
$\mu$ for almost all $\omega$ and all $s<t$;

(iii) if $\sum_i \lambda_i<0$, and $\lambda_i \neq 0$ for all $i$, then
$\mu_{\omega}$ is singular. \label{p.le}
\end{proposition}
A formula giving the dimension of $\mu_\omega$ is proved in \cite{Led+88}
under mild additional conditions.

The reliability of a single oscillator, {\em i.e.} that $\lambda < 0$,
is easily deduced from Prop.~\ref{p.le}: $\mu$ has a density because the
transition probabilities have densities, and no measure is preserved by
all the $F_{s,t,\omega}$ because different stimuli  distort the phase
space differently. Prop.~\ref{p.le}(i) and (ii) together imply that
$\lambda<0$.  See also, e.g., \cite{Pak+01,Nak+al05,Ter+03,Rit+03}.

The remarks above concerning $\mu$ apply also to the $2$-oscillator
model in Eq.~\eqref{2cell}. (That $\mu$ has a density is explained in
more detail in Sect. 3.2.) Therefore, Prop.~\ref{p.le}(i) and (ii)
together imply that $\lambda_1+\lambda_2 < 0$.  Here $\lambda_1=\lmax$
can be positive, zero, or negative.  If it is $>0$, then it will follow
from Prop.~\ref{p.le}(i) that $\lambda_2<0$, and by
Prop.~\ref{p.le}(iii), the $\mu_\omega$ are singular. From the geometry
of random SRB measures, we conclude that different initial conditions
are attracted to lower dimensional sets that depend on stimulus
history. Thus even in unreliable dynamics, the responses are highly
structured and far from uniformly distributed, as illustrated in
Fig.~\ref{f.snapshot}(b).

Finally, we observe that since $\lmax$ is nonrandom, the reliability of
a system is independent of the realization $\omega$ once the stimulus
amplitude $\eps$ is fixed.



\section{Coupling geometry and zero-input dynamics}

\subsection{Preliminary observations}

\begin{figure}
\begin{center}
\begin{multicols}{2}
\includegraphics[bb=0in 0in 2.92in 2.5in]{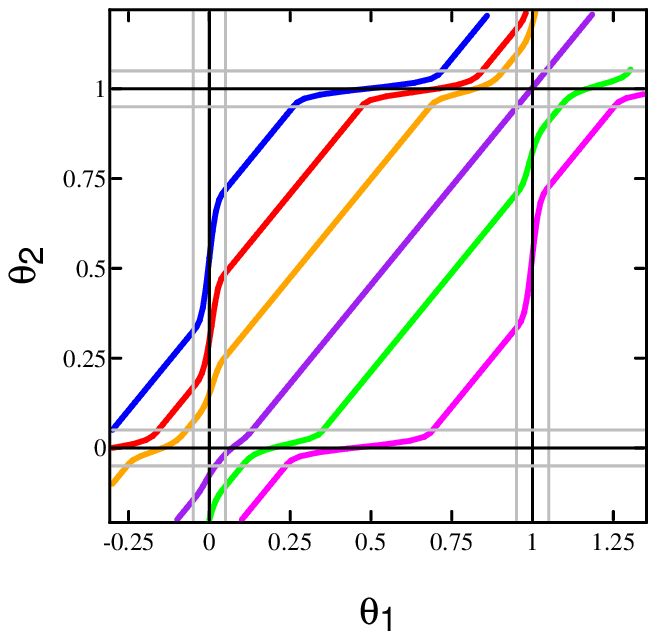}\\
(a) $\aff = 1, \afb = 1.5$

\columnbreak
\includegraphics[bb=0in 0in 2.86in 2.5in]{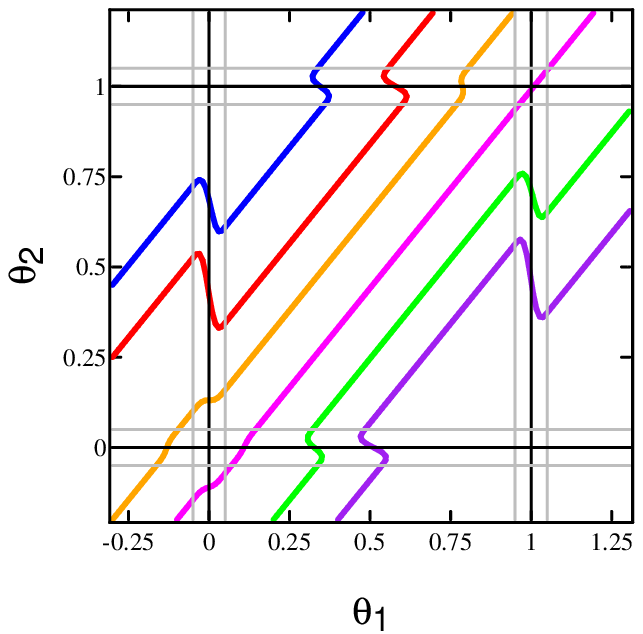}\\
(b) $\aff = -1, \afb = -0.5$
\end{multicols}
\end{center}
\caption{Plots of a few orbits of Eq.~(\ref{2cell}) with $\eps=0$ showing
 the strips in which flowlines are
  distorted.  In both sets, $\omega_1=1, \omega_2=1.1$. Note the
  directions of bending in relation to the signs of $\aff$ and $\afb$.}
\label{f.unforcedflow}
\end{figure}

First we describe the flow $\varphi_t$ on the 2-torus $\T^2$ defined by
Eq.~(\ref{2cell}) when the stimulus is absent, {\it i.e.}, when $\eps
=0$. We begin with the case where the two oscillators are uncoupled,
{\em i.e.} $\aff=\afb=0$. In this special case, $\varphi_t$ is a linear
flow; depending on whether $\omega_1/\omega_2$ is rational, it is
either periodic or quasiperiodic. Adding coupling distorts flow lines
inside the two strips $\{|\theta_1|<b\}$ and $\{|\theta_2|<b\}$.
These two strips correspond to the
regions where one of the oscillators ``spikes,'' transmitting a coupling
impulse to the other (see Sect. 1.1). For example, if $\aff>0$, then an
orbit entering the vertical strip containing $\theta_1=0$ will bend upward.
Because of the variable sensitivity of the receiving oscillator, the
amount of bending depends on $\aff$ as well as the value of $\theta_2$,
with the effect being maximal near $\theta_2=\frac{1}{2}$ and
negligible near $\theta_2=0$ due to our choice of the function
$z(\theta)$. The flow remains linear outside of these two strips. See
Fig.~\ref{f.unforcedflow}.

Because the phase space is a torus and Eq.~(\ref{2cell})
does not have fixed points for the parameters of interest, $\varphi_t$
has a global section, {\it e.g.}, $\Sigma_0=\{\theta_2=0\}$, with
a return map $T_0: \Sigma_0 \to \Sigma_0$. The
large-time dynamics of $\varphi_t$ are therefore described by
iterating $T_0$, a circle diffeomorphism. From standard theory,
we know that circle maps have Lyapunov exponents $\le 0$ for almost
all initial conditions with respect to Lebesgue measure. This in turn
translates into $\lmax =0$ for the flow $\varphi_t$.

\subsection{Reliability of two special configurations}
\label{geometric structure of pure ffw and fb}

We consider next two special but instructive cases of
Eq.~(\ref{2cell}), namely the ``pure feedforward'' case corresponding
to $\afb=0$ and the ``pure feedback'' case corresponding to $\aff=0$.
We will show that in these two special cases, the geometry of the system
prohibits it from being unreliable:

\begin{proposition}\label{prop.ff} For every $\eps > 0$,
\begin{itemize}
\item the system~(\ref{2cell}) has an ergodic stationary measure
$\mu$ with a density;
\item  (i) if $\afb=0$, then $\lmax\leq 0$;

(ii) if $\aff=0$, then $\lmax= 0$.
\end{itemize}
\end{proposition}

We first discuss these results informally.  Consider the case $\afb=0$.
Notice that when $\eps=0$, the time-$t$ map of the flow generated by
Eq.~(\ref{2cell}) sends vertical circles (meaning sets of the form
$\{\theta_1= c\}$ where $c$ is a constant) to vertical circles. As our
stimulus  acts purely in the
horizontal direction and its magnitude is independent of $\theta_2$,
vertical circles continue to be preserved by the flow-maps
$F_{s,t,\omega}$ when $\eps>0$.  
(One can also see this from the variational equations.) As is well known, $\lmax
> 0$ usually results from repeated stretching and folding of the
phase space. Maps that preserve all vertical circles are far too rigid
to allow this to happen. Thus we expect $\lmax \leq 0$ when $\afb=0$.
 In the pure feedback case $\aff=0$, the second oscillator rotates
freely without input from either external stimuli or the first
oscillator. Thus the system always has a zero Lyapunov exponent
corresponding to the uniform motion in the direction of
$\tht_2$.

Before proving Proposition~\ref{prop.ff}, we recall
some properties of Lyapunov exponents. Let $F_{0,t;\omega}$
denote the stochastic
flow on ${\mathbb T}^2$, and let $\mu$ be a stationary measure
of the process. Recall that for a.e. $\omega$, $\mu$-a.e.
$x \in {\mathbb T}^2$
and every tangent vector $v$ at $x$, the Lyapunov exponent
$$
\lambda_\omega(x,v) \ = \ \lim_{t \to \infty} \ \frac{1}{t} \log
|DF_{0,t;\omega}(x)v|
$$
is well defined. Moreover, if $\lambda_1 \le \lambda_2$ are the
Lyapunov exponents at $x$, then
\begin{equation}
  \lim _{t \to \infty} \ \frac{1}{t}  \log|\det(DF_{0,t;\omega}(x))| \ = \
  \lambda_1 + \lambda_2\ .
\label{det}
\end{equation}
We say $\{v_1, v_2\}$ is a {\it Lyapunov basis} at $x$
 if $\lambda_\omega(x,v_i)=\lambda_i$. It follows from
(\ref{det}) that for such a pair of vectors,
   \beq
    \lim_{t \to \infty} \ \frac{1}{t}  \log |\sin \angle(DF_{0,t;\omega}(x)v_1,
DF_{0,t;\omega}(x)v_2)| = 0 ,  \label{e.sineq}
    \eeq
that is, angles between images of vectors in a Lyapunov basis
do not decrease
exponentially fast. It is a standard fact that Lyapunov bases exist
almost everywhere, and that any nonzero tangent vector can be
part of such a basis.

\begin{proof}[Proof of Proposition~\ref{prop.ff}]
First, we verify that Eq.~(\ref{2cell}) has a stationary measure with a
density. Because the white noise stimulus $\eps I(t)$ instantaneously
spreads trajectories in the horizontal ($\tht_1$) direction, an
invariant measure must have a density in this direction. At the same
time, the deterministic part of the flow carries this density forward
through all of ${\mathbb{T}}^2$ since flowlines make approximately 45
degree angles with the horizontal axis. Therefore the two-oscillator
system has a 2-D invariant density whenever $\eps>0$.

We discuss the pure feedforward case. The case of $\aff=0$ is similar
and is left to the reader.

As noted above, when $\afb=0$ the flow $F_{0,t;\omega}$ leaves invariant
the family of vertical circles.  This means that $F_{0,t;\omega}$ {\it
  factors onto a stochastic flow on the first circle}. More precisely,
if $\pi : {\mathbb{T}}^2 = {\mathbb{S}}^1 \times {\mathbb{S}}^1 \to
{\mathbb{S}}^1$ is projection onto the first factor, then for almost
every $\omega$ and every $s<t$, there is a diffeomorphism $\bar
F_{0,t;\omega}$ of ${\mathbb{S}}^1$ with the property that for all $x
\in {\mathbb{T}}^2$, $\pi(F_{0,t;\omega}(x))= \bar F_{0,t;\omega}(\pi
(x))$. One checks easily that $\bar F_{s,t;\omega}$ is in fact the
stochastic flow corresponding to the system in which oscillator 2 is
absent and oscillator 1 alone is driven by the stimulus.  For
simplicity of notation, for $x \in {\mathbb{T}}^2$ and tangent vector
$v$ at $x$, we denote $\pi(x)$ by $\bar x$ and $D\pi(x)\cdot v$ by $\bar v$.
Furthermore, we let $\bar \lambda_\omega$ denote the
Lyapunov exponent  of $\bar F_{s,t;\omega}$.  From Prop.~\ref{p.le}, it
follows that $\bar \lambda_\omega <0$ almost everywhere.

We now consider the Lyapunov exponents of system (\ref{2cell}).  At
$\mu$-a.e. $x \in {\mathbb{T}}^2$,
we pick a vector $v$ in the $\theta_2$-direction. Since $F_{0,t;\omega}$ preserves vertical circles
and $\mu$ has a density, an argument similar to that before
Prop.~\ref{p.le} gives $\lambda_\omega(x,v) \le 0$.
Let $\{u,v\}$ be a Lyapunov basis at $x$. From
Eq.~(\ref{e.sineq}) and the relation between $F_{0,t;\omega}$, and
$\bar F_{s,t;\omega}$, we deduce that
$$
\lambda_\omega(x,u) = \lim_{t \to \infty} \ \frac{1}{t} \log
|\overline{DF_{0,t;\omega}(x)\cdot u}| = \lim_{t \to \infty} \
\frac{1}{t} \log |{D\bar{F}_{0,t;\omega}(\bar{x})\cdot\bar{u}}| = \bar
\lambda_\omega(\bar x, \bar u) < 0\ .
$$
\end{proof}

It is likely that $\lmax$ is typically $<0$ in the case $\afb=0$.  We
have shown that one of the Lyapunov exponents, namely the one
corresponding to the first oscillator alone, is $<0$.  The value of
$\lmax$ therefore hinges on $\lambda_\omega(x,v)$ for vectors $v$
without a component in the $\theta_1$-direction.  Even though this
growth rate also involves compositions of random circle maps, the maps
here are not {\it i.i.d.}: the kicks received by the second oscillator
are in randomly timed pulses that cannot be put into the form of the
white noise term in Eq.~(\ref{eq:sde}). We know of no mathematical
result that guarantees a strictly negative Lyapunov exponent in this
setting, but believe it is unlikely that Eq.~(\ref{eq:sde}) will have a
robust zero Lyapunov exponent unless $\aff=0$.

To summarize, we have shown that without recurrent connections, it is
impossible for a two-oscillator system to exhibit an unreliable
response.

\subsection{Phase locking in zero-input dynamics}

\begin{figure}
\begin{center}
\includegraphics[bb=0in 0in 2.7in 2.5in]{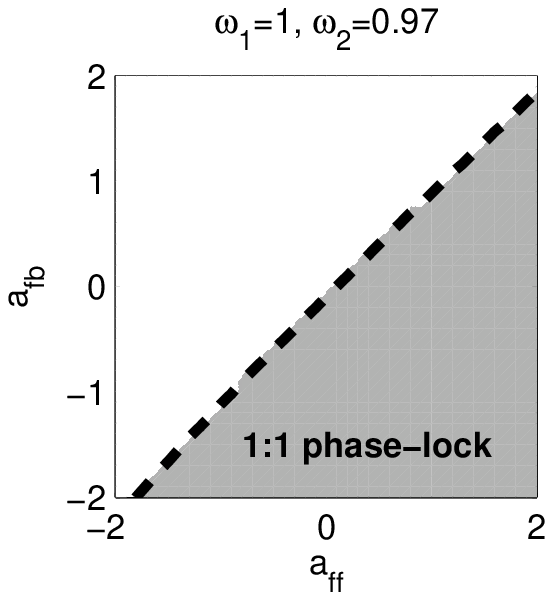}\qquad%
\includegraphics[bb=0in 0in 2.7in 2.5in]{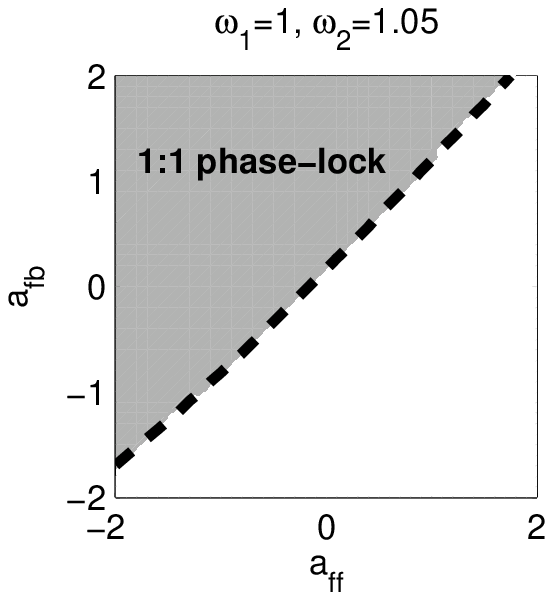}
\end{center}
\caption{The critical value $\afb^*$ as functions of $\aff$.  In both
  plots, the dashed line is the $\afb^*$ curve (see text), and the
  shaded regions are the parameters for which 1:1 phase-locking occurs.}
\label{f.afb-crit}
\end{figure}

This subsection is about the dynamics of the coupled oscillator system
when $\eps=0$. Recall that the intrinsic frequencies of the two
oscillators, $\omega_1$ and $\omega_2$, are assumed to be $\approx 1$.
Our main result is that if $\omega_1$ and $\omega_2$ differ by a
little, then in regions of the $(\aff,\afb)$-parameter plane in the form
of a square centered at $(0,0)$, {\it the two oscillators are 1:1
phase-locked for about half of the parameters}. Two examples are
shown in Fig.~\ref{f.afb-crit}, and more detail will be given as we
provide a mathematical explanation for this phenomenon.
(Phase locking of pairs of coupled phase oscillators is studied in
{\it e.g.}, \cite{gerstner96what,H95,erme96,Enm,CCC98,TH98}.
A primary difference is that we treat the problem on an open region
of the $(\aff,\afb)$-plane centered at $(0,0)$.)

\medskip
Let $\varphi_t$ be the flow on ${\mathbb T}^2$
defined by (\ref{2cell}) with $\eps =0$.  To study $\varphi_t$, we
consider its return map $T:\Sigma_b \to \Sigma_b$
where $\Sigma_b= \{\theta_2=b\}$.
Working with the section $\Sigma_b$ (as opposed to {\it e.g.} $\Sigma_0$)
simplifies the analysis as we will see, and substantively
the results are not section dependent.
 Let $\rho(T)$ be the rotation number of $T$.
Since $\omega_1 \approx \omega_2$, it is natural
to define $\rho(T)$ in such a way that $\rho(T)\approx 1$
when $\aff=\afb=0$, so that $\rho(T)$ may be interpreted as the
average number of rotations made by the first oscillator for each
rotation made by the second.  It is well known that if $\rho(T)$ is
irrational, then $\varphi_t$ is equivalent to a quasi-periodic flow via
a continuous change of coordinates, while $\rho(T) \in {\mathbb Q}$
corresponds to the presence of periodic orbits for $\varphi_t$.  In
particular, attracting fixed points of $T$ correspond to attracting
periodic orbits of $\varphi_t$ that are 1:1 phase-locked,
``1:1'' here referring to the fact that oscillator 2 goes around
exactly once for each rotation made by oscillator 1.

We begin with the following elementary facts:

\begin{lemma}

\begin{itemize}

\item With $\omega_1$, $\omega_2$, and $\aff$ fixed and letting
$T=T_{\afb}$, the function $\afb \mapsto T_{\afb} (\theta_1)$ is strictly
increasing for each $\theta_1$; it follows that $\rho(T_{\afb})$ is a
nondecreasing function\footnote{This is to allow for phase locking at
rational values of $\rho(T_{\afb})$.} of $\afb$.

\item With $\omega_1$, $\aff$, and $\afb$ fixed, $\omega_2 \mapsto
T_{\omega_2} (\theta_1)$ is strictly decreasing for each $\theta_1$, and
$\rho(T_{\omega_2})$ is a non-increasing function of $\omega_2$.

\end{itemize}
\noindent Analogous results hold as we vary $\aff$ and $\omega_1$
separately keeping  the other three quantities fixed.
\label{l.rotat}
\end{lemma}

\begin{proof}  This follows from the way each
coupling term bends trajectories; see Sect. 3.1 and
Fig.~\ref{f.unforcedflow}.  We show (a); the rest are proved similarly.
Consider two trajectories, both starting from the same point on $\Sigma_b$
but with different $\afb$. They will remain identical until their
$\tht_2$-coordinates
reach $1-b$, as $\afb$ does not affect this part of the flow. Now
at each point in the horizontal strip $H=\{1-b < \theta_2 < 1+b\}$,
the vector field corresponding to the larger
$\afb$ has a larger horizontal component while the vertical components
of the two vector fields are identical. It follows that the
trajectory with the larger $\afb$ will be bent farther to the right
as it crosses $H$.
\end{proof}

Our main result identifies regions in parameter space
where $T$ has a fixed point, corresponding to 1:1 phase-locking as
discussed above. We begin with the following remarks and notation:
(i) Observe that when $\omega_1=\omega_2$ and $\aff=\afb$,
we have ${T}(x)=x$ for all $x \in \Sigma_b$; this is a consequence
of the symmetry of
$z(\theta)$ about $\theta=\frac{1}{2}$. (ii) We introduce the notation $\Delta \omega = 1-\frac{\omega_1}{\omega_2}$, so that when
$\aff=\afb=0$, $x - T(x)=\Delta \omega$ for all $x$, {\it i.e.},
$\Delta \omega$ measures the distance moved by the
return map $T$ under the linear flow when the two oscillators are
uncoupled. Here, we have used $T$ to denote not only the section map
but its lift to $\mathbb R$ with $T(1)=1$ in a harmless abuse of
notation.

We will state our results for a bounded range of parameters.  For
definiteness, we consider $\aff, \afb \in [-2,2]$ and $.9 \le \om_i \le
1.1$. The bounds for $\aff$ and $\afb$ are quite arbitrary. Once chosen,
however, they will be fixed; in particular, the constants in our lemmas
below will depend on these bounds. {\it It is implicitly assumed that
  all parameters considered in Theorem 2 are in this admissible range}.
We do not view $b$ as a parameter in the same way as $\om_1, \om_2,
\aff$ or $\afb$, but instead of fixing it at $\frac{1}{20}$, we will
take $b$ as small as need be, and assume $|g|={\mathcal
  O}(\frac{1}{b})$ in the rigorous results to follow.

\begin{theorem}\label{T.plock} The following hold for all
admissible $(\om_1, \om_2, \aff, \afb)$ and all $b$ sufficiently small:

\begin{itemize}

\item If $\omega_2 > \omega_1$, then there exist
$\afb^*=\afb^*(\omega_1, \omega_2, \aff) >\aff$ and $\ell=\ell(\delom)>0$
such that $T$ has a fixed point for $\afb \in [\afb^*, \afb^*+\ell]$
and no fixed point for
$\afb<\afb^*$.

\item If $\omega_2 < \omega_1$, then there exist
  $\afb^*=\afb^*(\omega_1, \omega_2, \aff)<\aff$  and $\ell=\ell(\delom)>0$
  such that $T$ has a fixed point for $\afb \in [\afb^*-\ell, \afb^*]$ and no fixed point for $\afb>\afb^*$.
\end{itemize}

\noindent Moreover, $|\afb^*-\aff|= {\mathcal O}(\delom)$; and
for each $\delom \ne 0$, $\ell$ increases as $b$ decreases.

\end{theorem}

To prove this result, we need two lemmas, the proofs of both of which are given
in the Appendix. Define $\delafb=\afb-\aff$.

\begin{lemma} \label{l.T1}  There exist $b_1$ and $K>0$
such that for all admissible $(\om_1, \om_2, \aff, \afb)$, if
$b< b_1$ and $\delafb < Kb^{-2} \delom$, then $T(1-b)<1-b$.
\end{lemma}

\begin{lemma} \label{l.Thalf2}  There
exist $b_2, C>0$ and $x_1 \in (0, 1-b)$ such that for all admissible
$(\om_1, \om_2, \aff, \afb)$, if $b<b_2$ and $\delafb > C \delom$, then
$T(x_1)>x_1$.
\end{lemma}

\begin{proof}[Proof of Theorem~\ref{T.plock}.]
We prove (a); the proof of (b) is analogous.
Let $\om_1, \om_2$ and $\aff$ be given. Requirements on the size of
$b$ will emerge  in the course of the proof.

Observe first that with $\om_2>\omega_1$,  $T$ has
no fixed point (and hence there is no 1:1 phase locking) when $\afb=\aff$.
This is because
$T(x)=x$ when $\omega_2=\omega_1$ and $\afb=\aff$ as noted earlier,
and using Lemma~\ref{l.rotat}(b), we see that for
$\omega_2>\omega_1$ and $\afb = \aff $, the graph of ${T}$ is strictly
below the diagonal.

Keeping $\omega_1, \omega_2$ and $\aff$ fixed, we now increase $\afb$
starting from $\afb=\aff$.  By Lemma~\ref{l.rotat}(a), this
causes the graph of ${T}$ to shift up pointwise.  As $\afb$ is
gradually increased, we let $\afb^*$ be the first parameter at which
the graph of ${T}$ intersects the diagonal, {\em i.e.} where
there exists $x^*\in\Sigma_b$ such that ${T}(x^*)=x^*$ --- if
such a parameter exists. Appealing once more to Lemma~\ref{l.rotat},
we see that $\rho(T)<1$ for all $\afb < \afb^*$, so $T$ can have no
fixed point for these values of $\afb$.

We show now that $\afb^*$ exists, and that the phase-locking persists
on an interval of $\afb$ beyond $\afb^*$. First, if $b$ is small enough,
then by Lemma~\ref{l.T1}, $T(1-b)<1-b$ for all $\afb < \aff +
Kb^{-2} \delom$. Now if $b$ is small enough that
$Kb^{-2} > C$ where $C$ is as in Lemma~\ref{l.Thalf2}, then
for $\afb \in [\aff + C\delom, \aff + Kb^{-2} \delom]$, $T(x_1)>x_1$
for some $x_1< 1-b$.  For $\afb$ in this range,  $T$ maps
the interval $[x_1, 1-b]$ into itself, guaranteeing a fixed point.
It follows that (i) $\afb^*$ exists and is $<\aff +C\delom$, and
(ii) $T$ has a fixed point for an interval of $\afb$ of length
$\ell \ge (Kb^{-2} - C)\delom$. This completes the proof.
\end{proof}

As noted in the proof, for as long as $\om_2 \ne \om_1$, the lengths of the
phase-locked intervals, $\ell$, can be made arbitrarily large by taking
$b$ small. On the other hand, if we fix $b$ and
shrink $\delom$, then these intervals will shrink. This is consistent with
the phenomenon that the phase-locked region lies on opposite sides
of the diagonal $\afb=\aff$ when we decrease $\omega_2$ past
$\omega_1$, as shown in Fig.~\ref{f.afb-crit}.

Instead of tracking the numerical constants in the proofs, we have
checked numerically that for $b=\frac{1}{20}$, the pictures
in Fig.~\ref{f.afb-crit} are quite typical, meaning about $50\%$ of
the parameters are phase locked.
Specifically, for $\delom$ up to about $10\%$ and $|\aff|<2$,
$\afb^*-\aff \lesssim 0.2$, so that the $\afb^*$-curve describing
the onset of phase-locking is still quite close to the diagonal $\aff=\afb$.
Also, for $\delom$ as small as $\frac12 \%$, the phase locked intervals
have length $\ell>4$. These facts together imply that for parameters
in the admissible range, the pictures are as shown in Fig.~\ref{f.afb-crit}.

\medskip

\begin{figure}
\begin{center}
\includegraphics[bb=0in 0in 2.5in 2in]{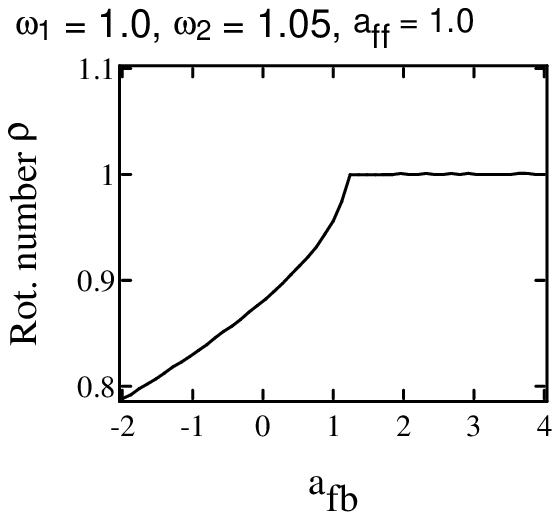}\qquad%
\includegraphics[bb=0in 0in 2.5in 2in]{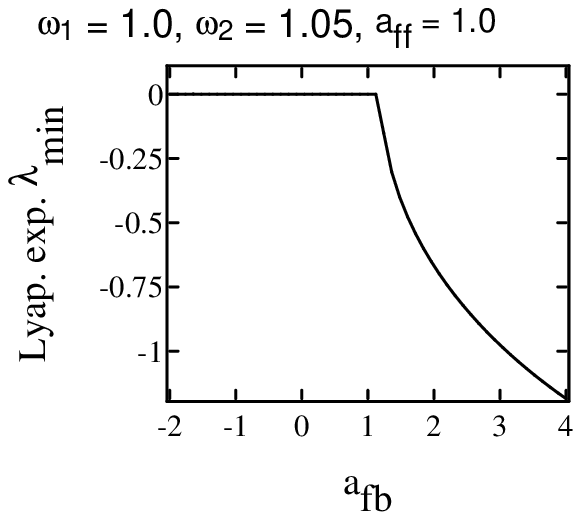}
\end{center}
\caption{The rotation number $\rho$ and Lyapunov exponent
  $\lmin$ as functions of $\afb$.}
\label{phaselock-lyap-rn}
\end{figure}


Fig.~\ref{phaselock-lyap-rn} shows numerically-computed rotation
numbers $\rho$ and the rates of contraction to the corresponding limit
cycle, {\em i.e.} the smaller Lyapunov exponent $\lmin$ of the flow
$\varphi_t$.
Notice that as $\afb$ increases past $\afb^*$, $\lmin$ decreases
rapidly, so that the fixed point of $T$ becomes very stable, a fact
consistent with the large interval on which the system is 1:1
phase-locked. Furthermore, for the full range of $|\aff|, |\afb|
\leq 2$ and $0.9 \le \om_i \le 1.1$, we find numerically that $0.53 <
\rho(T) < 1.89$.  Phase-locking corresponding to rational $\rho(T)$
with small denominators $q$ ({\it e.g.}, $q=3,4,5$) is detected, but
the intervals are very short and their lengths decrease rapidly with
$q$. In other words, when the system is not
1:1 phase-locked -- which occurs for about $50\%$ of the parameters of
interest -- modulo fine details the system appears to be roughly
quasi-periodic over not too large timescales. When the white-noise
stimulus $\eps I(t)$ is added, these fine details will matter little.


\section{Reliable and Unreliable Behavior}
\label{reliable-and-unreliable}

Numerical evidence is presented in Sect.~\ref{shear in 2-osc} showing
that unreliability can occur even when the stimulus amplitude is
relatively small, and that its occurrence is closely connected with the
onset of phase-locking in the zero-input system.  A geometric
explanation in terms of {\em shear-induced chaos} is proposed.
Additionally, other results leading to a qualitative understanding of
$\lmax$ as a function of $\aff, \afb$ and $\eps$ are discussed in
Sect.~\ref{futher observations}.

\subsection{A brief review of shear-induced chaos}
\label{s.shearreview}

\begin{figure}
\begin{center}
\includegraphics[bb=0 0 307 117,scale=0.85]{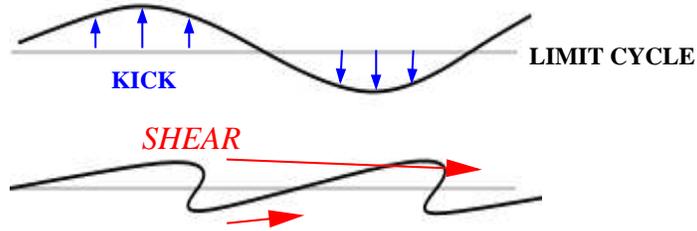}
\end{center}
\caption{The stretch-and-fold action of a kick followed by
  relaxation in the presence of shear.}
\label{f.shear}
\end{figure}

A rough idea of what we mean by ``shear-induced chaos'' is depicted in
Fig.~\ref{f.shear}: An external force is transiently applied to a limit
cycle (horizontal line), causing some phase points to move up and some
to move down. Suppose the speeds with which points go around the limit
cycle are height dependent. If the velocity gradient, which we refer to
as {\it ``shear'',} is steep enough, then the bumps created by the
forcing are exaggerated as the system relaxes, possibly causing the
periodic orbit to fold. Such folding has the potential to lead to the
formation of strange attractors.  If the damping is large relative to
the velocity gradient or the perturbation size, however, the bumps in
Fig.~\ref{f.shear} will dissipate before any significant stretching
occurs.

This subsection reviews a number of ideas surrounding the mechanism
above.  This mechanism is known to occur in many different dynamical
settings. We have elected to introduce the ideas in the context of {\em
  discrete-time kicking of limit cycles} instead of the setting of
Eq.~(\ref{2cell}) because the geometry of discrete-time kicks is more
transparent, and many of the results have been shown numerically to
carry over with relatively minor modifications.  Extensions to relevant
settings are discussed later on in this subsection. A part of this body
of work is supported by rigorous analysis. Specifically, theorems on
shear-induced chaos for periodic kicks of limit cycles are proved in
\cite{WY1,WY2,WY3,WY4}; it is from these articles that many of the
ideas reviewed here have originated. Numerical studies extending the
ideas in \cite{WY2,WY3} to other types of underlying dynamics and
forcing are carried out in \cite{ly07}. For readers who wish to see a
more in-depth (but not too technical) account of the material in this
subsection, \cite{ly07} is a good place to start.  In particular, Sect.
1 in \cite{ly07} contains a fairly detailed exposition of the rigorous
work in~\cite{WY1,WY2,WY3,WY4}.

\bigskip
\noindent {\bf Discrete-time kicks of limit cycles}

\medskip
We consider a flow $\varphi_t$ in any dimension, with a limit cycle
$\gamma$. Let $T_0<T_1<T_2< \cdots$ be a sequence of kick times, and
let $\kappa_n, n=0,1,2, \cdots$, be a sequence of kick maps (for the
moment $\kappa_n$ can be  any transformation of the phase space). We
consider a system kicked at time $T_n$ by $\kappa_n$, with ample time
to relax between kicks, {\it i.e.}, $T_{n+1}-T_n$ should not be too
small on average.

Central to the geometry of shear-induced chaos is
the following dynamical structure of the unforced system:
For each $x \in
\gamma$, the {\it strong stable manifold} $W^{ss}(x)$ of $\varphi_t$
 through $x$ is defined to be
$$
W^{ss}(x) =\{y:\lim_{t\to\infty}d(\varphi_t(x),\varphi_t(y))=0\}\ .$$
These codimension 1 submanifolds are invariant under the flow, meaning
$\varphi_t$ carries $W^{ss}(x)$ to $W^{ss}(\varphi_t(x))$. In
particular, if $\tau$ is the period of the limit cycle, then
$\varphi_\tau(W^{ss}(x))=W^{ss}(x)$ for each $x$. Together these
manifolds partition the basin of attraction of $\gamma$ into
hypersurfaces, forming what is called the {\it strong stable foliation}.

\begin{figure}
\begin{center}
\includegraphics*[bb=72 0 371 103,scale=0.9]{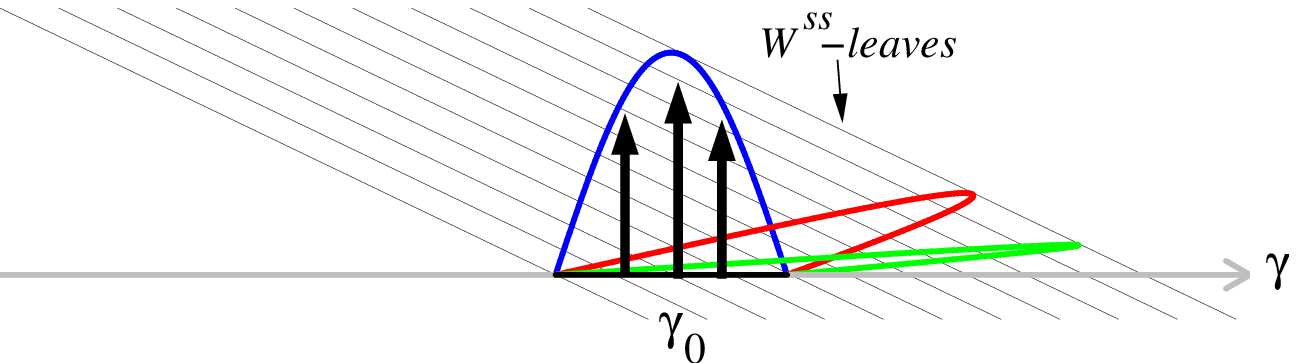}
\end{center}
\caption{Geometry of folding in relation to the $W^{ss}$-foliation.
  Shown are a segment $\gamma_0\subset\gamma$, its image after one kick,
  and two of the subsequent images under the flow.}
\label{f.wss}
\end{figure}

Fig.~\ref{f.wss} shows a segment $\gamma_0 \subset \gamma$, its image
$\gamma_0^+=\kappa(\gamma)$ under a kick map $\kappa$, and two images
of $\gamma_0^+$ under $\varphi_{n\tau}$ and $\varphi_{m\tau}$ for $n>m
\in {\mathbb Z}^+$.  If we consider integer multiples of $\tau$, so
that the flow-map carries each $W^{ss}$-leaf to itself, we may think of
it as sliding points in $\gamma_0^+$ toward $\gamma$ along
$W^{ss}$-leaves. (For $t$ that are not integer multiples of $\tau$, the
picture is similar but shifted along $\gamma$.)  The stretching created
in this combined kick-and-slide action depends both on the geometry of
the $W^{ss}$-foliation and on the action of the kick. Fig.~\ref{f.wss}
sheds light on the types of kicks that are likely to lead to folding: A
forcing that drives points in a direction roughly parallel to the
$W^{ss}$-leaves will not produce folding.  Nor will kicks that
essentially carry one $W^{ss}$-leaf to another, because such kicks
necessarily preserve the ordering of the leaves.
 What causes the stretching and folding is the {\it
variation} along $\gamma$ in how far points $x\in\gamma$ are kicked as
measured in the direction transverse to the $W^{ss}$-leaves; we think
of this as the ``effective deformation'' of the limit cycle $\gamma$ by
the kick.

To develop a quantitative understanding of the factors conducive to the
production of chaos, it is illuminating to consider the following
linear shear model~\cite{Z,WY2}:
\begin{equation}
\begin{array}{rcl}
\dot{\theta} & =& 1 + \sigma y\ ,\\
\dot{y} &=& -\lambda y + A \sin(2 \pi \theta)
\cdot\sum_{n=0}^\infty \delta(t-nT)\ .
\end{array}
\label{sine shear flow}
\end{equation}
Here, $\theta \in {\mathbb{S}}^1$ and $y \in {\mathbb R}$ are phase
variables, and $\sigma, \lambda, A$ are constants.\footnote{In this
subsection, $\lambda$ denotes the damping constant in Eq.~(\ref{sine
    shear flow}).}  We assume for definiteness that $\sigma$ and $\lambda$
are $>0$, so that when $A=0$, the system has a limit cycle $\gamma$ at
$\{y=0\}$.  Its $W^{ss}$-leaves are easily computed to be straight lines
with slope $=-\frac{\lambda}{\sigma}$.  When $A \ne 0$, the system is
kicked periodically with period $T$. For this system, it has been shown
that the {\it shear ratio}
\begin{equation}
\frac{\sigma}{\lambda}\cdot A \ \equiv \ \frac{\mbox{shear}}{\mbox{damping}} \cdot \mbox{deformation} \ = \ \frac{1}{|\mbox{slope}(W^{ss})|}
\cdot \mbox{deformation}
\label{shear ratio}
\end{equation}
is an excellent predictor of the dynamics of the system. Roughly speaking,
if $\lmax$ denotes the largest observed Lyapunov exponent, then

\medskip
\begin{itemize}
\item if the shear ratio is sufficiently large, $\lmax$ is likely to be $>0$;
\item if the shear ratio is very small, then $\lmax$ is slightly
$<0$ or equal to $0$;
\item as the shear ratio increases from small to large, $\lmax$ first becomes negative, then becomes quite irregular
  (taking on both positive and negative values), and is eventually
  dominated by positive values.
\end{itemize}

\medskip
\noindent To get an idea of why this should be the case, consider the
composite kick-and-slide action in Fig.~\ref{f.wss} in the context of
Eq.~(\ref{sine shear flow}). The time-$T$ map of Eq.~(\ref{sine shear
  flow}) is easily computed to be
\begin{equation}
\begin{array}{rcl}
\theta_T & = & \theta + T + \frac{\sigma}{\lambda}\cdot\big[
  A\sin(2\pi\theta) + y\big] \cdot \big(1 - e^{-\lambda
  T}\big)\qquad\mbox{(mod 1)}\ ,\vspace{1.5ex}\\
y_T & = & e^{-\lambda T}\cdot\big(y + A\sin(2\pi\theta)\big)\ .
\end{array}
\label{linear shear flow time-T map}
\end{equation}
When the contraction in $y$ is sufficiently strong, the first component of
this map gives a good indication of what happens in the full $2$-D system.
As an approximation, define $f_T(\theta)=\theta_T$ and view $f_T$ as
a map of $\gamma$ to itself. When the shear ratio
is large and $(1 - e^{-\lambda T})$ is not too small, $|f_T'|$ is
quite large over much of $\gamma$, and the associated expansion
has the potential to create the positive $\lmax$ mentioned in (a).
At the opposite extreme, when the shear ratio is very small,
$f_T$ is a perturbation of the identity; this is the scenario in (b).
Interestingly, it is for intermediate shear ratios that $f_T$ tends to have
sinks, resulting in $\lmax<0$ for the $2$-D system.
The $1$-D map $f = \lim_{T \to \infty} f_T$ is known
variously as the {\em phase resetting curve} or the {\em singular limit}.
It is used heavily in~\cite{WY1,WY2,WY3,WY4} to produce rigorous
results for large $T$.

We now return to the the more general setting of $\varphi_t$
with discrete
kicks $\kappa_n$, and try to interpret the results above as best we can.  To
make a meaningful comparison with the linear shear flow, we propose to
first put our unforced system in ``canonical coordinates,'' {\it i.e.},
to reparametrize the periodic orbit $\gamma$ so it has unit speed, and
to make the kick directions perpendicular to $\gamma$ ---
assuming the kicks have well defined directions.
In these new coordinates, sizes of vertical deformations make sense,
as do the idea of damping and shear, even
though these quantities and the angles between $W^{ss}$-manifolds
and $\gamma$ all vary along $\gamma$.  Time intervals between
kicks may vary as well.  The general geometric picture is thus
a distorted version of the linear shear flow. We do not believe there is
a simple formula to take the place of the shear ratio in this general
setting; replacing the quantities $\sigma, \lambda$ and $A$ by their averages
is not quite the right thing to do. We emphasize, however, that while system details affect the numerical values of $\lmax$ and the amount of shear needed
to produce $\lmax>0$, the fact that {\it the overall trends
as described in (a)--(c) above are valid} has been repeatedly demonstrated in
simulation; see, {\em e.g.},~\cite{ly07}.

\bigskip
\newpage
\noindent {\bf Generalization to stochastic forcing}

\medskip We now replace the discrete-time kicks in the discussion above
by a directed (degenerate) continuous stochastic forcing, {\it i.e.} by
a term of the form $V(x)dW_t$ where $V$ is a vector field and $dW_t$ is
white noise.  By Trotter's product formula, the dynamics of the
resulting stochastic flow can be approximated by a sequence of composite
maps of the form $\varphi_{\Delta t} \circ \kappa_{\Delta t, \omega}$
where $\Delta t$ is a small time step, $\kappa_{\Delta t, \omega}$ are
kick maps (of random sizes) in the direction of $V$, and
$\varphi_{\Delta t}$ is the unforced flow.  Most of the time, the maps
$\kappa_{\Delta t, \omega}$ have negligible effects.  This is especially
the case if the size of $V$ is not too large and damping is
present. Once in a while, however, a large deviation occurs, producing
an effect similar to that of the discrete-time kicks at times $T_n$
described above.  Cast in this light, we expect that the ideas above
will continue to apply -- albeit without the factors in the shear ratio
being precisely defined.

We mention two of the differences between stochastic forcing and
periodic, discrete kicks.  Not surprisingly, stochastic forcing gives
simpler dependence on parameters: $\lmax$ varies smoothly,
irregularities of the type in (c) above having been ``averaged
out''. Overall trends such as those in (a)--(c) tend to be unambiguous
and more easily detected than for deterministic kicks.  Second, unlike
periodic kicks, very small forcing amplitudes can elicit chaotic
behavior without $\frac{\sigma}{\lambda}$ being very large; this is
attributed to the effects of large deviations.

\bigskip
\noindent {\bf Generalization to quasi-periodic flows}

\medskip We have chosen to first introduce the ideas above in the
context of limit cycles where the relevant geometric objects or
quantities (such as $\sigma, \lambda$ and $W^{ss}$) are more easily
extracted.  These ideas apply in fact to flows on a torus that are
roughly ``quasi-periodic'' -- meaning that orbits may or may not be
periodic but if they are, the periods are large -- provided the
forcing, stochastic or discrete, has a well-defined direction as
discussed earlier.  The main difference between the quasi-periodic
setting and that of a limit cycle is that $W^{ss}$-leaves are generally
not defined. A crucial observation made in \cite{ly07} is that since
folding occurs in finite time, what is relevant to the geometry of
folding is not the usual $W^{ss}$-foliation (which takes into
consideration the dynamics as $t \to \infty$) but {\it finite-time}
strong stable foliations. Roughly speaking, a time-$t$ $W^{ss}$-leaf is
a curve segment (or submanifold) that contracts in the first $t$ units
of time. For the ideas above to apply, we must verify that time-$t$
$W^{ss}$-manifolds exist, have the characteristics of a large shear
ratio, and that $t$ is large enough for the folding to actually occur.  If these
conditions are met, then one can expect shear-induced chaos
to be present for the same reasons as before.


\subsection{Phase-locking and unreliability in the two-cell
  model}
\label{shear in 2-osc}

\begin{figure}
\begin{center}
\includegraphicsbw{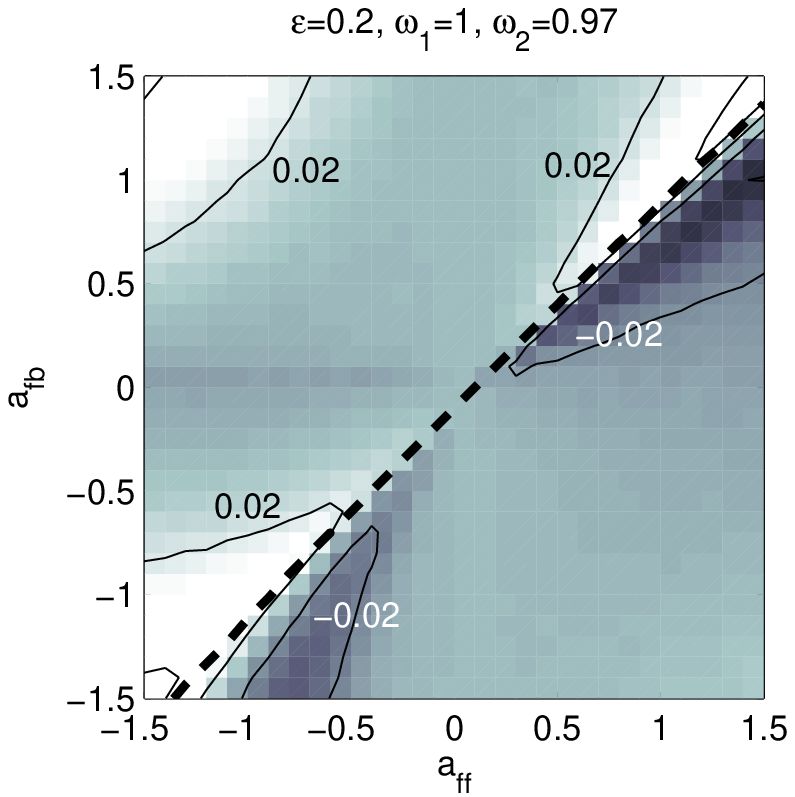}{bb=0in 0in 3.4in 3.2in}%
\hspace*{-18pt}\includegraphicsbw{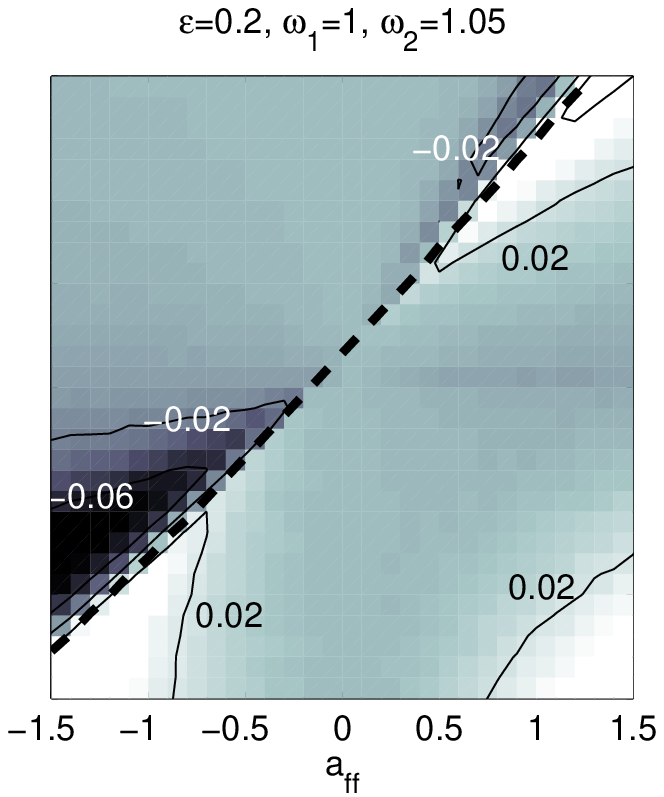}{bb=0in 0in 3in 3.2in}%
\hspace*{-18pt}\includegraphicsbw{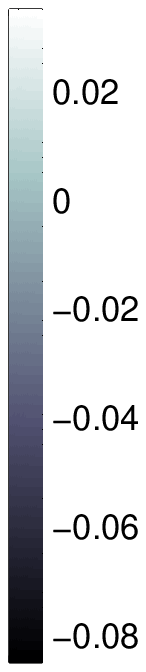}{bb=0in 0in 0.5in 3.2in}\\
(a) $\eps=0.2$\\\vspace{24pt}
\includegraphicsbw{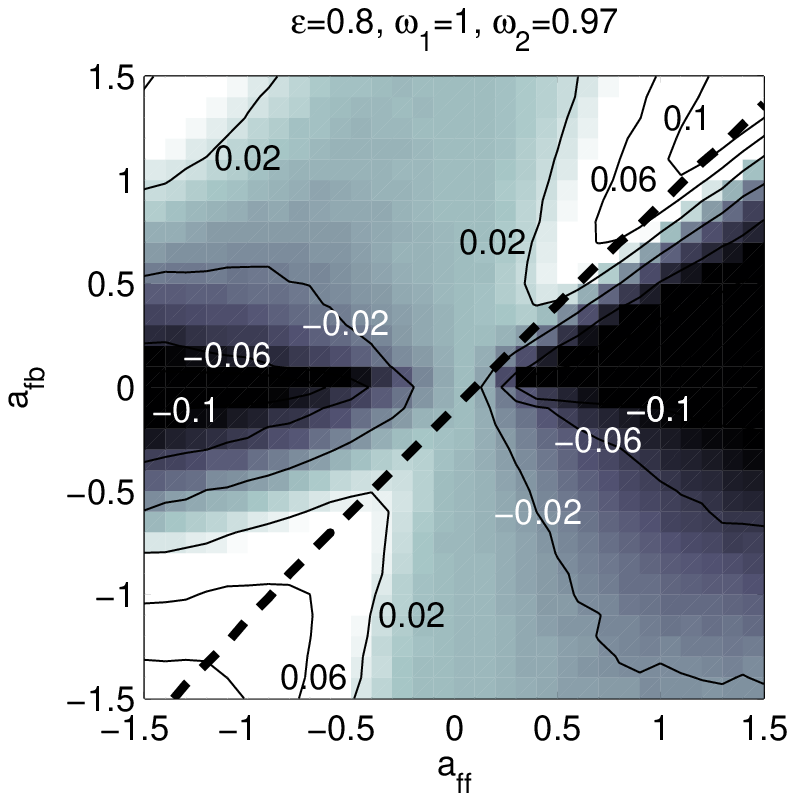}{bb=0in 0in 3.4in 3.2in}%
\hspace*{-18pt}\includegraphicsbw{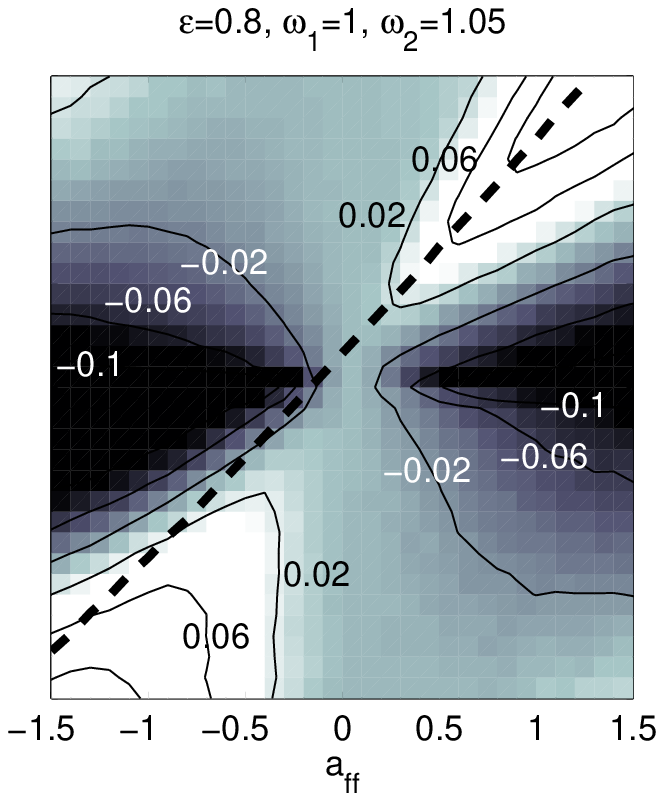}{bb=0in 0in 3in 3.2in}%
\hspace*{-18pt}\includegraphicsbw{graphics/LE_aff_afb_colorbar}{bb=0in 0in 0.5in 3.2in}\\
(b) $\eps=0.8$
\end{center}
\caption{Maximum Lyapunov exponent $\lmax$ versus coupling strengths in
  the two-cell network.  In all plots, we use $\omega_1=1$.  The
  $\afb^*$-curve from Fig.~\ref{f.afb-crit} is overlaid.  All computed
  $\lmax$ shown here have standard errors of $\leq 0.002$ as estimated
  by the method of batched means.  By the Central Limit Theorem, this
  means the actual $\lmax$ should lie within $\approx 2.5\times 0.002 =
  0.005$ of the computed value with $\gtrsim 99\%$ probability.  {\bf
    Remark on plots:} We have chosen the dynamic range in shading the
  figures to allow meaningful comparison of figures; a side effect is
  that some contour lines may not be visible.  We always indicate the
  actual range of values through explicit labels.  }
\label{f.LEaffafb}
\end{figure}

We now return to reliability questions of Eq.~(\ref{2cell}).  In this
subsection and the next, numerical data on $\lmax$ as functions of
$\aff, \afb$ and $\eps$ are discussed.  Our aim is to identify the
prominent features in these numerical results, and to propose
explanations for the phenomena observed.

The present subsection is limited to phenomena related to the onset of
phase-locking, which we have shown in Sect. 3.3 to occur at a curve in
$(\aff,\afb)$-space that runs roughly along the diagonal. We will refer
to this curve as the $\afb^*$-curve. Fig.~\ref{f.LEaffafb} shows $\lmax$
as a function of $\aff$ and $\afb$ for several choices of parameters,
with $\afb^*$-curves from Fig.~\ref{f.afb-crit} overlaid for ease of
reference. In the top two panels, where $\eps$ is very small, evidence
of events connected with the onset of phase-locking is undeniable:
definitively reliable ($\lmax<0$) and definitively unreliable
($\lmax>0$) regions are both present.  Continuing to focus on
neighborhoods of the $\afb^*$-curves, we notice by comparing the top and
bottom panels that for each $(\aff, \afb)$, the tendency is to shift in
the direction of unreliability as $\eps$ is increased. We will argue in
the paragraphs to follow that these observations are entirely consistent
with predictions from Sect. 4.1.

\bigskip
\noindent {\bf Shearing mechanisms}

\smallskip
For concreteness, we consider the case $\omega_2>\omega_1$,
and consider first parameters at which the unforced system has a limit
cycle, {\it i.e.}, for each $\aff \in [-1.5,1.5]$, we consider values of $\afb$
that are $> \afb^*$ and not too far from $\afb^*$.  From Sect. 4.1,
we learn that to determine the propensity of the system for
shear-induced chaos, we need information on (i) the geometry of
the limit cycle, (ii) the orientation of its $W^{ss}$-manifolds in relation
to the cycle, and (iii) the effective deformation due to the forcing.

The answer to (i) is simple: As with all other trajectories, the limit
cycle is linear with slope $ \gtrsim1$ outside of the two corridors
$|\theta_1|<b$ and $|\theta_2|<b$, where it is bent; see
Fig.~\ref{f.unforcedflow}. As for (ii) and (iii), we already know what
happens in two special cases, namely when $\aff=0$ or $\afb=0$. As
discussed in Sect. 3.2, when $\afb=0$, vertical circles are invariant
under $\varphi_t$. Since $W^{ss}$-leaves are the only manifolds that
are invariant, that means the $W^{ss}$-manifolds are vertical. We noted
also that the forcing preserves these manifolds. In the language of
Sect. 4.1, this means the forcing creates {\it no variation}
transversal to $W^{ss}$-leaves:  the ordering of points in this
direction is preserved under the forcing.
Hence shear-induced chaos is not possible here, and not likely for
nearby parameters. A similar argument (which we leave to the reader)
applies to the case $\aff=0$. From here on, we assume $\aff, \afb$ are
both away from $0$.

\begin{figure}
\begin{center}
\resizebox{3.2in}{!}{\includegraphics*{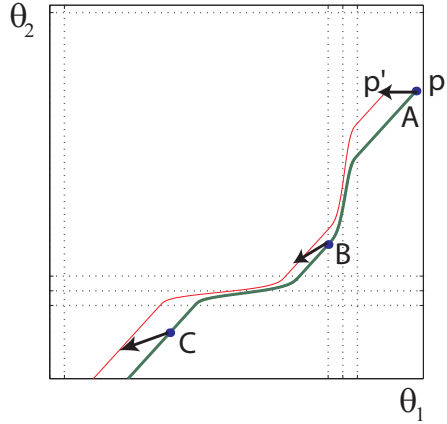}}
\end{center}
\caption{Rotation of vectors in {\it backwards} time.  Here,
  $\omega_1=1, \omega_2=1.1, \aff=1, \afb=1.5$.}
  \label{f.sstable}
\end{figure}

We now turn to a treatment of (ii) when
$\aff, \afb$ are both away from $0$, and claim that $W^{ss}$-leaves
generally have a roughly diagonal orientation, {\it i.e.}, they point
in a roughly southwest-northeast (SW-NE) direction.
To find this orientation, we fix a point $p$ on the limit cycle
$\gamma$, and any nonzero tangent vector $v$ at $p$ (see
Fig.~\ref{f.sstable}). We then flow backwards in time, letting
$p_{-t}=\varphi_{-t}(p)$ and $v_{-t}=D\varphi_{-t}(p)v$. The strong
stable direction at $p_{-t}$ is the limiting direction of
$v_{-t-n\tau}$ ($\tau =$ period of $\gamma$) as $n\to \infty$. Exact
orientations of $W^{ss}$-leaves depend on $\aff,\afb$
and are easily computed numerically.

The following simple analysis demonstrates how to deduce the general
orientation of the $W^{ss}$-leaves in the two-oscillator system from
the signs of its couplings. For definiteness, we consider $\aff>0$, so
that $\afb^*$ is also positive and slightly larger than $\aff$. Here, a
typical situation is that if we identify the phase space with
the square $[0,1] \times [0,1]$, then the limit cycle crosses
the right edge $\{1\} \times [0,1]$ in the bottom half,  and the bottom edge
$[0,1] \times \{0\}$ in the right half (as shown in Fig.~\ref{f.sstable}).
Let $A,B$ and $C$ be as shown, and consider a
point $p$ at $A$. Flowing backwards, suppose it takes time $t_B$ to
reach point $B$, and time $t_C$ to reach point $C$. We discuss how
$v_{-t}$ changes as we go from $A$ to $C$. The rest of the time the
flow is linear and $v_{-t}$ is unchanged.

\smallskip
\noindent {\it From $A$ to $B$:} Compare the backward orbits of $p$ and
$p'$, where $p'=p+k v$ and $k>0$ is thought of as infinitesimally
small. As these orbits reach the vertical strip $\{g>0\}$, both are
bent downwards due to $\aff>0$.  However, the orbit of $p'$ is bent
more because the function $z(\theta)$ peaks at $\theta=1/2$ (see
Fig.~\ref{f.sstable}).  Thus, $v_{-t_B} = v + (0, -\delta_1)$ for some
$\delta_1>0$.

\smallskip
\noindent {\it From $B$ to $C$:} Continuing to flow
backwards, we see by an analogous argument that as the two
orbits cross the horizontal strip $\{g>0\}$, both are bent
to the left, and the orbit of $p'$ is bent more.  Therefore,
$v_{-t_C}=v_{-t_B} + (-\delta_2, 0)$ for some $\delta_2>0$.

\smallskip Combining these two steps, we see that each time
the orbit of $p_{-t}$ goes from $A$ to $C$, a vector of the form
$(-\delta_2, -\delta_1)$  is added to $v_{-t}$. We conclude that as $t
\to \infty$, the direction of $v_{-t}$ is asymptotically SW-NE.
Moreover, it is a little more W than S compared to the limit cycle
because $v_{-t}$ must remain on the same side of the cycle.

\smallskip

\begin{figure}
\begin{center}
\begin{tabular}{cp{12pt}c}
\includegraphics[bb=0 4 223 203,scale=0.92]{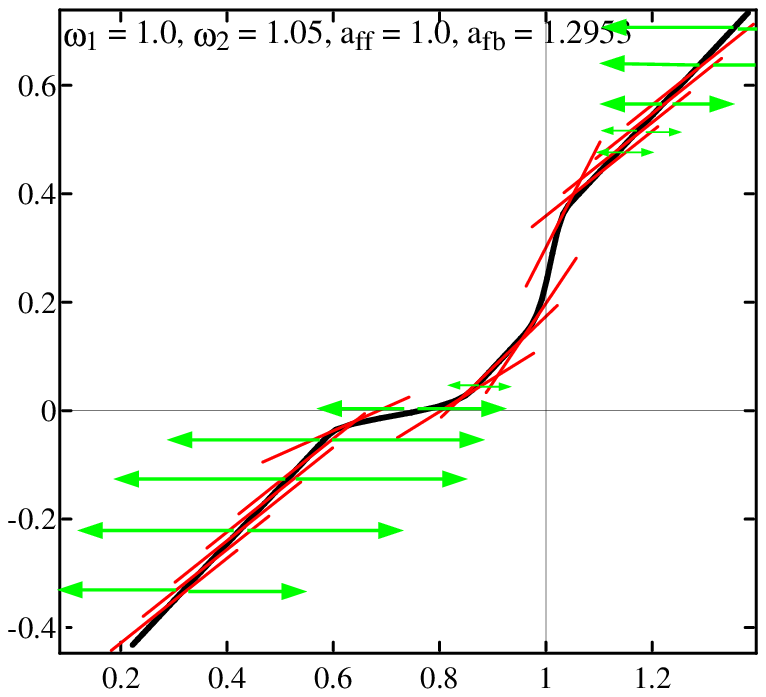}&&
\includegraphics[bb=0in 0in 2.60in 2.75in,scale=0.92]{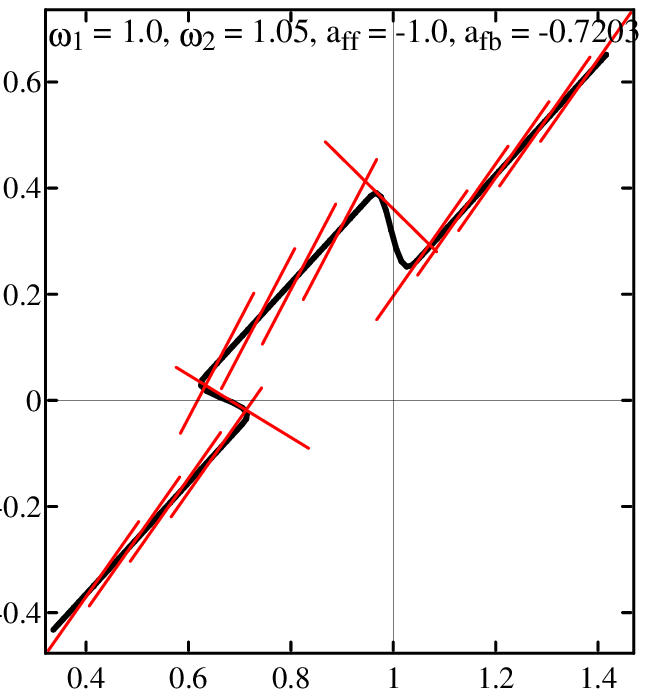}\\
(a) $\aff,\afb>0$ && (b) $\aff,\afb<0$\\
\end{tabular}
\end{center}
\caption{Strong stable directions along limit cycles.  In (a),
  $\afb=\afb^*+0.1$; in (b), $\afb=\afb^*+0.2$. Additionally in (a), horizontal
  lines with arrows indicate the impact of the forcing at various points
  along the cycle; the variable impact is due to the
  $z$-function.}
\label{comparing shear}
\end{figure}

Numerical computations of strong stable directions are shown in
Fig.~\ref{comparing shear}. The plot in (a) is an example of the
situation just discussed. The plot in (b) is for $\aff, \afb<0$, for
which a similar analysis  can be carried out. Notice how small the
angles are between the limit cycle and its $W^{ss}$-leaves; this is
true for all the parameter sets  we have examined where $\afb$ is
close to $\afb^*$. Recall
from Sect. 4.1 that it is, in fact, the angles in {\it canonical
coordinates} that count. Since the limit cycle is roughly diagonal and
the forcing is horizontal, putting the system in canonical coordinates
will not change these angles by more than a
moderate factor (except in one small region in picture (a));
{\it i.e.}, the angles will remain small.

Finally, we come to (iii), the deformation due to the forcing.
Given that the forcing is in the horizontal
direction and its amplitude depends on $\theta_1$ (it is negligible
when $\theta_1\approx 0$ and has maximal effect when
$\theta_1\approx\frac{1}{2}$),  it causes ``bump-like''
perturbations transversal to the $W^{ss}$-manifolds
(which are roughly SW-NE) with a geometry similar to that in
Fig.~\ref{f.wss}; see Fig.~\ref{comparing shear}(a).

This completes our discussion of the limit cycle case. We move now to
the other side of the $\afb^*$-curve, where the system is, for
practical purposes, quasi-periodic (but not far from periodic). As
discussed in the last part of Sect. 4.1, the ideas of shear-induced
chaos continue to apply, with the role of $W^{ss}$-leaves now played by
finite-time stable manifolds. Since these manifolds change slowly with
$\aff$ and $\afb$, it can be expected -- and we have checked -- that
they continue to make small angles with flowlines. Likewise, the
forcing continues to deform flowlines by variable amounts as measured
in distances transversal to finite-time stable manifolds.

\medskip We conclude that when $\aff, \afb$ are both away from $0$, the
geometry is favorable for shear-induced stretching and folding.
Exactly how large a forcing amplitude is needed to produce
a positive $\lmax$ depends on system details. Such information
cannot be deduced from the ideas reviewed in Sect.~4.1 alone.

\bigskip
\noindent {\bf Reliability--unreliability interpretations}

\smallskip
We now examine more closely Fig.~\ref{f.LEaffafb}, and attempt to
explain the reliability properties of those systems whose couplings lie
in a neighborhood of the $\afb^*$-curve. The discussion
below applies to $|\aff|>$ about $0.3$. We have observed earlier that
for $\aff$ or $\afb$ too close to $0$, phase-space geometry prohibits
 unreliability.

Consider first Fig.~\ref{f.LEaffafb}(a), where the stimulus amplitude
$\eps=0.2$ is very weak. Regions showing positive and negative Lyapunov
exponents are clearly visible in both panels. Which side of the $\afb^*$-curve
corresponds to the phase-locked region is also readily recognizable to
the trained eye (lower triangular region in the picture on the left and
upper triangular region on the right; see Fig.~4).

We first explore the phase-locked side of the $\afb^*$-curve. Moving
away from this curve, $\lmax$ first becomes definitively negative. This
is consistent with the increased damping noted in Sect. 3.2; see
Fig.~6(b). As we move farther away from the $\afb^*$-curve still,
$\lmax$ increases and remains for a large region close to $0$.
Intuitively, this is due to the fact that for these parameters the
limit cycle is very robust. The damping is so strong that the forcing
cannot (usually) push any part of the limit cycle any distance away
from it before it is brought back to its original position. That is to
say, the perturbations are negligible. With regard to the theory in
Sect. 4.1, assuming $\sigma$ remains roughly constant, that $\lmax$
should increase from negative to $0$ as we continue to move away from
$\afb^*$ is consistent with increased damping; see (b) and (c) in the
interpretation of the shear ratio.

Moving now to the other side of the $\afb^*$-curve, which is essentially quasi-periodic, regions of
unreliability are clearly visible. These regions in fact begin slightly
on the phase-locked side of the curve, where a weakly attractive limit
cycle is present. We have presented evidence to support our contention
that this is due to shear-induced chaos, or folding of the phase space.
The fact that $\lmax$ is more positive before the limit cycle is born
than after can be attributed to the weaker-to-nonexistent damping
before its birth. Thus the general progression of $\lmax$ from roughly
$0$ to definitively negative to positive as we cross the $\afb^*$-curve
from the phase-locked side to the quasi-periodic side is altogether
consistent with scenarios (a)--(c) in Sect. 4.1 together with the
observations in the paragraph on stochastic forcing.

We point out that the unreliability seen in these panels is fairly
delicate, perhaps even unexpected {\it a priori} for the smaller values
of $\aff$ and $\afb$, such as $0.3 < |\aff|, |\afb| < 0.8$:
The bending of the flow lines is rather mild at these smaller coupling parameters.
Moreover, we know that no chaotic
behavior is possible at $\eps=0$, and the stimulus amplitude of
$\eps=0.2$ in the top panels is quite small. Recall, however, that
the stimulus is a fluctuating white noise, and $\eps$ gives only
an indication of its {\it average} amplitude. As noted in Sect.~4.1, we believe the
unreliability seen is brought about by an interaction between
the large fluctuations in the stimulus presented and the shearing
in the underlying dynamics.

In Fig.~\ref{f.LEaffafb}(b), the stimulus amplitude is increased to
$\eps=0.8$. A close examination of the plots shows that near to and on both
sides of the $\afb^*$-curve,  $\lmax$ has increased for each parameter
pair $(\aff, \afb)$, and that the reliable regions are pushed deeper
into the phase-locked side compared to the top panels. This is
consistent with the shear ratio increasing with forcing amplitude as
predicted in Sect. 4.1.

This completes our discussion in relation to Fig.~\ref{f.LEaffafb}.

\medskip

\begin{figure}
\includegraphics[bb=0in 0in 1.55in 1.55in]{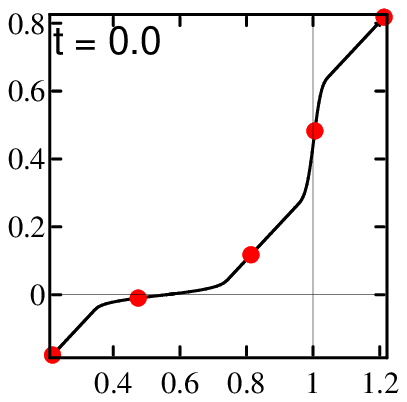}\hspace*{\fill}%
\includegraphics[bb=0in 0in 1.56in 1.55in]{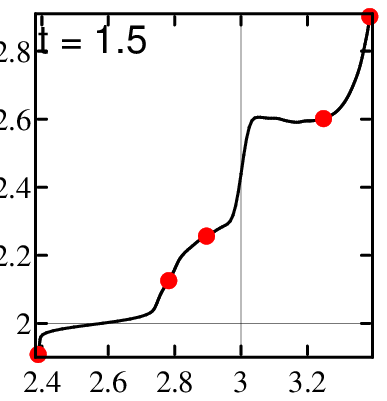}\hspace*{\fill}%
\includegraphics[bb=0in 0in 1.55in 1.55in]{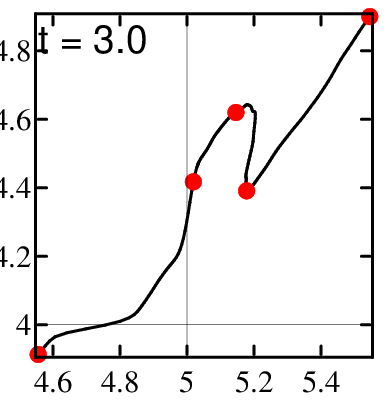}\hspace*{\fill}%
\includegraphics[bb=0in 0in 1.57in 1.55in]{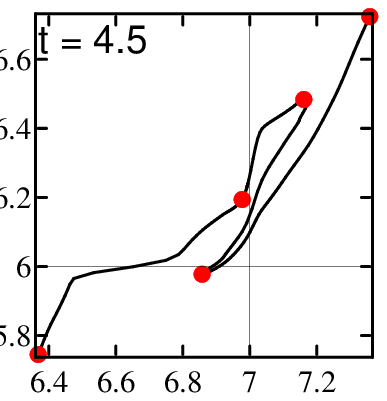}\\
\caption{Folding action caused by white noise forcing and shear near the
  limit cycle (with $\afb>\afb^*$).  At $t=0$, the curve shown is the
  lift of the limit cycle $\gamma$ to $\R^2$.  The remaining panels show
  lifts of the images $F_{0,t,\omega}(\gamma)$ at increasing times.  The parameters are $\omega_1=1$,
  $\omega_2=1.05$, $\aff=1$, $\afb=1.2$, and $\eps=0.8$.  Note that it
  is not difficult to find such a fold in simulations: very roughly, 1 out of
  4 realizations of forcing gives such a sequence for $t\in[0,5]$. }
\label{f.foldingcycle}
\end{figure}

To complement the theoretical description of the geometry of folding
given in Sect. 4.1, we believe it is instructive to see an actual
instance of how such a fold is developed when the system (\ref{2cell})
is subjected to an arbitrary realization of white noise. A few
snapshots of the time evolution of the limit cycle under such a forcing
is shown in Fig.~\ref{f.foldingcycle}. Notice that at the beginning,
the combined action of the coupling and forcing causes the curve to
wriggle left and right in an uncertain manner, but once a definitive
kink is developed (such as at $t=3$), it is stretched by the shear as
predicted.

\subsection{Further observations on parameter dependence}
\label{futher observations}

\begin{figure}[t]
\begin{center}
\includegraphicsbw{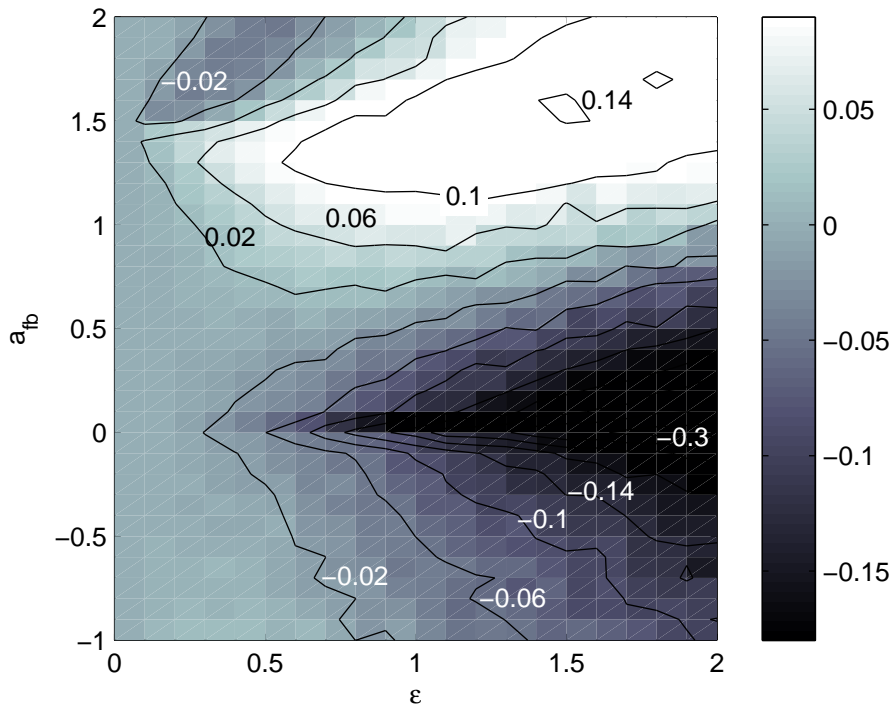}{bb=0in 0in 5in 4in}
\end{center}
\caption{$\la_{\max}$ as function
  of $\afb$ and $\eps$, with $\omega_1=1$, $\omega_2=1.1$,
  and $\aff=1$.}
\label{f.triplept}
\end{figure}

We now proceed to other observations on the dependence of $\lmax$
on network parameters. Our aim is to identify the salient features
in the reliability profile of the oscillator system (\ref{2cell}) as a function of
$\aff, \afb$ and $\eps$, and to attempt to explain the phenomena
observed. Our observations are based on plots of the type in
Figs.~\ref{f.LEaffafb} and \ref{f.triplept}. Some of the explanations
we venture below are partial and/or speculative; they will be so indicated.

\begin{enumerate}

\item {\it Triple point:} This phenomenon is the main topic of
  discussion in Sect. 4.2. Fig.~\ref{f.triplept} shows a different view
  of the parameter dependence of $\lmax$.  At about $\afb=1.4$, which is near $\afb^*$ for the parameters
  used, both positive and negative Lyapunov exponents are clearly
  visible for very small $\eps$ in a manner that is consistent with
  Fig.~\ref{f.LEaffafb} (even though the $\omega$'s differ slightly from that figure).
  We note again that this region, which we refer to as the ``triple point,'' is an area of extreme sensitivity
  for the system, in the sense that the system may respond in a
  definitively reliable or definitively unreliable way to stimuli of
  very small amplitudes, with the reliability of its response depending
  sensitively on coupling parameters.

\item {\it Unreliability due to larger couplings:} Fig.~\ref{f.LEaffafb}
  and other plots not shown point to the occurrence of
  unreliability when $|\aff|$ and $|\afb|$ are both relatively large.
  We are referring here specifically to the ``off-diagonal'' regions of
  unreliability (far from the $\afb^*$-curve) in Fig.~\ref{f.LEaffafb}.
  This phenomenon may be partly responsible for the unreliability seen
  for the larger values of $\aff$ and $\afb$ on the diagonal as well; it
  is impossible to separate the effects of the different mechanisms.

  We do not have an explanation for why one should expect $\lmax>0$
  for larger $|\aff|$ and $|\afb|$ aside from the obvious, namely that
  tangent vectors are more strongly rotated as they cross the strips
  $\{|\theta_i|<b\}$, making it potentially easier for folding to occur.
  But folding does {\it not} occur at $\eps=0$ in spite of this larger bending.
 We believe the difference between the two situations is due to the following
  noise-assisted mechanism: For $p \in {\mathbb T}^2$ and a tangent
  vector $u$ at $p$, let us say $u$ is {\it positively oriented} with
  respect to flowlines if starting from the direction of the flow and
  rotating counter-clockwise, one reaches $u$ before reaching the
  direction of the backward vector field.  Without external forcing, if
  $u$ is positively oriented, $D\varphi_t(p)\cdot u$ will remain
  positively oriented for all $t$, because these vectors cannot cross
  flowlines.  Now, in order for folding to occur, as in the formation of a
  horseshoe, the flow-maps must reverse the orientations of
  {\em some} tangent vectors.
  Even though larger values of $|\aff|$ and $|\afb|$ mean that tangent
  vectors are more strongly rotated, a complete reversal in direction
   cannot be accomplished without crossing flowlines.  A small
  amount of noise makes this crossing feasible, opening the door
  (suddenly) to positive exponents.

\item {\it The effects of increasing $\eps$} (up to around
  $\eps=2$):\vspace{6pt}
\begin{itemize}

\item {\it Unreliable regions grow larger, and $\lmax$ increases:}
A natural explanation here is that the stronger the stimulus, the greater
  its capacity to deform and fold the phase space -- provided such
  folding is permitted by the underlying geometry.  Because of the form
  of our stimulus, however, too large an amplitude simply pushes all phase
  points toward $\{ \theta_1=0 \}$.  This will not lead to $\lmax>0$, a
  fact supported by numerics (not shown).\vspace{6pt}

\item {\it Reliable regions grow larger, and the responses become
   more reliable}: As $\eps$ increases, the reliable region
    includes all parameters $(\aff, \afb)$ in a large wedge
  containing the $\afb=0$ axis. Moreover, in this region $\lmax$ becomes
  significantly more negative as $\eps$ increases.  We propose the
  following heuristic explanation: In the case of a single oscillator,
  if we increase the amplitude of the stimulus, $\lmax$ becomes more
  negative. This is because larger distortions of phase space geometry
  lead to more uneven derivatives of the flow-maps $F_{s,t;\omega}$,
  which in turn leads to a larger gap in Jensen's Inequality (see the
  discussion before Prop. 2.1).  For two oscillators coupled as Eq.~(\ref{2cell}), increasing $\eps$ has a similar stabilizing effect on
  oscillator 1.  Feedback kicks from oscillator 2 may destabilize it, as
  happens for certain parameters near the $\afb^*$-curve.  However, if
  $\aff$ is large enough and enhanced by a large $\eps$, it appears that
  the stabilizing effects will prevail.

\end{itemize}


\end{enumerate}

\newpage
\section{Conclusions}

We have shown that a network of two pulse-coupled phase oscillators can
exhibit both reliable and unreliable responses to a white-noise stimulus,
depending on the signs and strengths of network connections and
the stimulus amplitude. Specifically:

\begin{enumerate}

\medskip
\item (a) {\it Dominantly feedforward} networks are always reliable,
and they become more reliable

\quad with increasing input amplitude.

\smallskip
(b) {\it Dominantly feedback} networks are always neutral to weakly reliable.

\medskip
\item When {\it feedback and feedforward coupling strengths are comparable},
whether both are negative (mutually inhibitory) or positive (mutually excitatory),
we have observed the following:

\smallskip
(a) For {\it weak input amplitudes}, the system is extremely sensitive to
coupling strengths, with

\quad substantially reliable and
unreliable configurations occurring in close proximity. This

\quad phenomenon is explained by mechanisms of phase locking and
shear-induced chaos.

\smallskip
(b) For {\it stronger input amplitudes}, the system is typically
unreliable.

\end{enumerate}

\medskip
\noindent For weak stimuli, the most reliable configurations
are, in fact, not dominantly feedforward but those with comparable
feedforward--feedback couplings.

\bigskip
We expect these results to be fairly prototypical for certain
pulse-coupled phase oscillators, such as those with ``Type I'' phase response curves that frequently occur
in neuroscience.  Understanding the effects of qualitatively
different coupling types or oscillator models is an interesting topic
for future study.

Another natural extension is to larger networks.  This is the topic of
the companion paper~\cite{Lin+al07_ppr2}, where we use the present
two-oscillator system as an embedded ``module'' to illustrate how
unreliable dynamics can be generated locally and propagated to other
parts of a network.



%
%

\bigskip \textbf{Acknowledgements:} K.L. and E.S-B. hold NSF Math. Sci.
Postdoctoral Fellowships and E.S-B. a Burroughs-Wellcome Fund Career
Award at the Scientific Interface; L-S.Y. is supported by a grant from
the NSF.  We thank Brent Doiron, Bruce Knight, Alex Reyes, John Rinzel,
and Anne-Marie Oswald for helpful discussions over the course of this
project.

\section*{APPENDIX:  Proof of Theorem 2}
\addcontentsline{toc}{section}{Appendix}
\label{lemmas}

Here we prove the two lemmas needed for Theorem 2.

Let $\om_1,\aff, \delom$ and $\delafb$ be given, with $\delom, \delafb>0$.
To study the system where $\om_2$ is defined by
$\delom=1-\frac{\om_1}{\om_2}$ and $\afb=\aff+\delafb$, we
will seek to compare trajectories for systems with the following
parameter sets:
    \beqn
    \begin{array}{lllll}
\mbox{System A} :  &  \aff, & \afb=\aff,   & \om_1, & \om_2 = \om_1  \\
\mbox{System B} : & \aff, & \afb=\aff & \om_1, & \om_2 = \om_1 + \delom\cdot \om_2 \\
\mbox{System C} : & \aff, & \afb=\aff+\delafb, & \om_1, & \om_2
= \om_1 + \delom\cdot \om_2
    \end{array}
    \eeqn
That is, System C is the system of interest, and Systems A and B are
used to help analyze System C. We introduce also the following notation:
$H$ and $V$ denote the horizontal and vertical strips of width $2b$
centered at integer values of $\theta_2$ and $\theta_1$.  We will
work in $\R^2$ instead of $\T^2$.

\bigskip

\noindent {\bf{Lemma~\ref{l.T1}:}} {\it There exist $b_1$ and $K>0$
such that for all admissible $(\om_1, \om_2, \aff, \afb)$, if
$b< b_1$ and $\delafb < Kb^{-2} \delom$, then $T(1-b)<1-b$.}

\begin{figure}
\begin{center}
\includegraphics[bb=131 360 468 618,clip,scale=0.95]{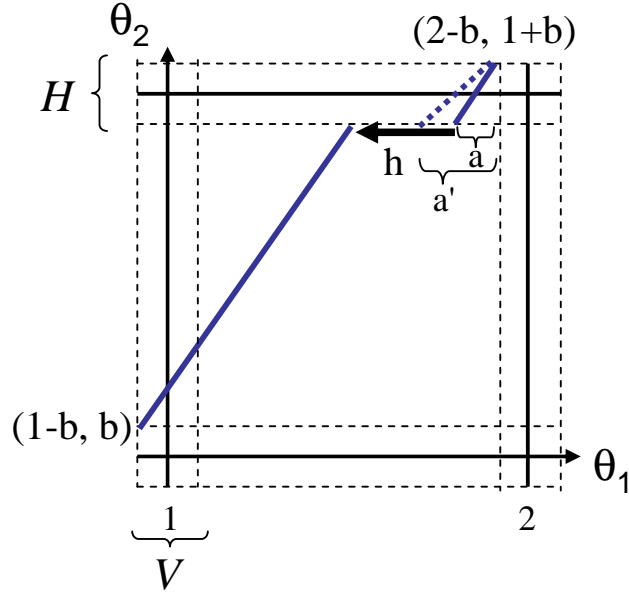}
\end{center}
\caption{Values used in proving Lemma~\ref{l.T1}.
}
\label{f.lemma-fig-1}
\end{figure}

\begin{proof} For each of the 3 parameter sets above, two orbits are considered:
Orbit 1 starts from $(\theta_1, \theta_2)=(1-b,b) \in \Sigma_b$
and runs forward in time until
it meets $\Sigma_{1-b}=\{\theta_2=1-b\}$; orbit 2
starts from $(2-b,1+b) \in \Sigma_{1+b}$ and runs backward in time until
it meets $\Sigma_{1-b}$. We need to prove that for System C,
under the conditions in the lemma, the end point of orbit 1 lies to the left
of the end point of orbit 2 (as shown in Fig.~\ref{f.lemma-fig-1}).
This is equivalent to $T(1-b)<1-b$.

For System A, orbits 1 and 2 meet, since
for this set of parameters, $T(x)=x$ for all $x$ as noted in Sect. 3.3.
Comparing Systems  A and B, since the vector
field for System B has greater slope everywhere, and outside of
$H \cup V$ it has slope $\frac{\om_2}{\om_1}>1$, we conclude that
for System B the end point of orbit 1 lies to the left
of the end point of orbit 2, with a separation $h > \delom/2$;
see Fig.~\ref{f.lemma-fig-1}.

Next, we compare Systems B and C. Orbit 1 for the two systems is
identical, since the equation outside of $H$ does not involve $\afb$.
Orbit 2, however, differs for the two systems. To estimate by how much,
we compare $a$, the distance from the end point of orbit 2 to
$\theta_1=2$ for System B, and $a'$, the corresponding distance for
System C as marked in Fig.~\ref{f.lemma-fig-1}.  First, there exist
$b_1$ and $k_1>0$ such that for $\theta_1 \in (2-5b_1, 2)$, we have
$z(\theta_1) < k_1(2-\theta_1)^2$ and $|z'(\theta_1)|<2k_1(2-\theta_1)$.
Shrinking $b_1$ further if necessary, we have that for $b<b_1$, orbit 2
has slope $> 1/2$ everywhere and therefore the entire orbit, for both
Systems B and C, lies within the region $H \cap \{2-5b<\theta_1< 2-b\}$.
The next step is to apply Gronwall's Lemma to a system that incorporates
both Systems B and C. Since $\theta_2(t)$ is identical for the two
systems in the relevant region, we may write the equations as
    \begin{eqnarray*}
     \dot{\tht_1} & = & - \om_1
- (\aff + \delta) z(\tht_1) \hat g(t) \nonumber \\
\dot{\delta}&=& 0 \; \;,
    \end{eqnarray*}
where $\delta=0$ corresponds to System B, $\delta =
\delafb$ corresponds to System C, and $\hat g(t)=g(\theta_2(t))$.
Notice that each trajectory reaches $\Sigma_{1-b}$ after a time
$\tau=2b/\om_2$.

Motivated by the observation that $z(\tht_1) = {\mathcal O}(b^2)$
in the relevant rectangle, we rescale the variable $\delta$
by $\bar \delta = b^2 \delta$, and define $\bar z(\tht_1) = \frac{1}{b^2} z(\tht_1)$.
This gives
    \begin{eqnarray}
     \dot{\tht_1} & = & - \om_1  \label{e.ODE1}
- (b^2 \aff + \bar \delta) \bar z(\tht_1) \hat g(t) \\
\dot{\bar \delta}&=& 0 \; \;. \label{e.ODE2}
    \end{eqnarray}
To find $a$, we solve \eqref{e.ODE1}-\eqref{e.ODE2} over the time
interval $[0,\tau]$, starting from $(\tht_1(0),\bar \delta(0))=(1-b,0)$;
to find $a'$, we do the same, starting from $(1-b,b^2
\delafb)$. Applying Gronwall's Lemma gives $|a' -a| < b^2 \delafb \exp
(L \tau)$, where $L$ is the Lipschitz constant for the allied vector
field.  To estimate $L$, note that $|\hat g|={\cal{O}}(\frac1b)$, $|\bar
z|={\cal{O}}(1)$, and $|\bar z'|={\cal{O}}(\frac1b)$.  This gives $L
={\cal{O}}(1/b)$, and $\exp (L \tau)={\cal{O}}(1)$.  Therefore, $|a' -a| <
k_2 b^2 \delafb$ for some constant $k_2$.  Note that this constant can
be made independent of $\om_1, \om_2, \aff$ or $\afb$.

Recall from the first part of the proof that to obtain the desired result
for System C, it suffices to guarantee $|a' -a| < h$. This happens when
    \begin{equation}
 \delafb < \frac{\delom}{2 k_2 b^2} := Kb^{-2} \delom  \; \; . \label{e.L1ineq}
    \end{equation}
\end{proof}


\noindent {\bf{Lemma~\ref{l.Thalf2}:}}  {\it There exist $b_2, C>0$
and $x_1 \in (0, 1-b)$
such that for all admissible $(\om_1, \om_2, \aff, \afb)$, if
$b<b_2$ and $\delafb > C \delom$, then $T(x_1)>x_1$.}

\begin{proof} All orbit segments considered in this proof run from
$\Sigma_b \cap \{\theta_1 \in (0,1)\}$ to $\Sigma_{1+b}$.  We assume
  $\aff > 0$; the case $\aff < 0$ is similar.  First, we fix $x_0, x_1>b$ so that for
  all admissible $(\om_1, \om_2, \aff, \afb)$, the trajectory starting
  from $x_1$ intersects $H=\{1-b<\theta_2<1+b\}$ in $H \cap \{\theta_1
  \in (1+x_0,\frac32)\}$.  Such $x_0, x_1$ clearly exist for small
  enough $b$.  Starting from $x_1$, we compare the trajectories for
  Systems A, B and C.  We know that the trajectory for System A will end
  in $(1+x_1,1+b)$. Thus to prove the lemma, we need to show the
  trajectory for System C ends to the right of this point.  This
  comparison is carried out in two steps:

\medskip
\noindent Step 1: Comparing Systems A and B.
We claim that the horizontal separation of the end points of these two
trajectories is $<c \delom$ for some constant $c>0$.
It is not a necessary assumption, but the comparison is simpler if
we assume $b$ is small:
First, a separation $c_1 \delom$ in the $\tht_2$ direction develops
between the trajectories as they flow linearly from $(x_1,b)$ to $\tht_1=(1-b)$. Next, while the trajectories flow through $V$,
Gronwall's lemma can be used in a manner similar to the above to show
they emerge from $V$ with a separation $\le c_2 \delom$ in the
$\tht_2$ direction. Third is the region of linear flow, resulting
in a separation $ \le c_3 \delom$ in the $\tht_1$ direction as the
trajectories enter $H$. Gronwall's lemma's is again used in the final stretch
as the trajectories traverse $H$.

\medskip
\noindent Step 2: Comparing Systems B and C.
Notice that up until they reach $\Sigma_{1-b}$, the two trajectories
are identical. In $H$, their $\tht_2$ coordinates are equal, and
the crossing time is $\tau=\frac{2b}{\om_2}$.
Let $\tht_1^B(t)$ and $\tht_1^C(t), \ t \in [0,\tau]$, denote their $\tht_1$ coordinates while in $H$. We write
$$
\tht_1^C(\tau) - \tht_1^B(\tau) \ = \
\int_0^\tau \aff (z(\theta_1^C(t)) - z(\tht_1^B(t))) g(\tht_2(t))dt \ + \ \int_0^\tau \delafb z(\theta_1^C(t)) g(\tht_2(t))dt\ .
$$
The first integral is $\ge 0$ by design: via our choice of $x_1$,
we have arranged to have $1<\tht^B_1(t)<\tht_1^C(t)<\frac32$, and
the $z$-function is monotonically increasing between
$\theta_1=1$ and $\theta_1=\frac32$. As for the second integral,
we know $z(\tht^C_1(t))$ is bounded away from $0$ since
$1+x_0< \tht^C_1(t) < \frac32$, so the integral is $> d \delafb$
for some constant $d>0$.
It follows that $T(x_1)>x_1$ if $d \delafb > c \delom$.
 \end{proof}


\bibliographystyle{plain}
\bibliography{main}

\end{document}